\newcommand{\Msun}{\mbox{$M_{\odot}$}}
\newcommand{\Rsun}{\mbox{$R_{\odot}$}}

\documentclass{aa}
\usepackage{graphicx}

\begin{document}
   \title{The Star Cluster Population of M51: II. Age distribution and
   relations among the derived parameters}

   \titlerunning{Stellar Clusters in M51: II.}

   \author{N. Bastian \inst{1,2}, M. Gieles \inst{1},
          H.J.G.L.M. Lamers \inst{1,3}, R.A. Scheepmaker \inst{1}, R. de Grijs \inst{4}
       }  
       
    \authorrunning{N. Bastian et al.}

   \offprints{bastian@astro.uu.nl}

   \institute{$^1$Astronomical Institute, Utrecht University, 
              Princetonplein 5, NL-3584 CC Utrecht, The Netherlands \\
              \email{bastian@astro.uu.nl} \\
              $^2$European Astronomical Institute, Karl-Schwarzchild-Strasse 2 
               D-85748 Garching b. M\"{u}nchen, Germany \\
             $^3$SRON Laboratory for Space Research, Sorbonnelaan 2, 
             NL-3584 CA Utrecht, The Netherlands\\
             $^4$Department of Physics \& Astronomy, University of
              Sheffield, Hicks Building, Hounsfield Road, Sheffield
              S3 7RH, UK  \\}

   \date{Received 14 April 2004 / Accepted 28 October 2004}


   \abstract{We use archival {\it Hubble Space Telescope} observations
   of broad-band 
   images from the ultraviolet (F255W-filter) through the near
   infrared (NICMOS F160W-filter) to study the star
   cluster population of the interacting spiral galaxy M51. We obtain
   age, mass, extinction, and effective radius 
   estimates for 1152 star clusters in a region of $\sim 7.3 \times
   8.1$ kpc centered on the nucleus and extending into the outer
   spiral arms.   In this paper we present the data set and
   exploit it to determine the age distribution and relationships
   among the fundamental parameters (i.e. age, mass, effective
   radius).  We show the critical dependence of the age distribution
   on the sample selection, and confirm that using a constant mass
   cut-off, above which the sample is complete for the entire age
   range of interest, is essential.  In particular, in this sample we
   are complete only for masses above 5$\times 10^{4}$ M$_{\odot}$
   for the last 1 Gyr.  Using this dataset we find: {\it i}) that the
   cluster formation rate seems to have had a large increase 
   $\sim$ 50-70 Myr ago, which is coincident with the suggested {\it
   second passage} of its companion, NGC 5195, {\it ii}) a large number
   of extremely young ($<$ 10 Myr) star clusters, which we interpret
   as a population of unbound clusters of which a large majority
   will disrupt within the 
   next $\sim$10 Myr, and {\it iii)} that the distribution of cluster sizes
   can be well approximated by a power-law with exponent, $ -\eta =
   -2.2 \pm 0.2$, which is very similar to that of Galactic globular clusters,
   indicating that cluster disruption is largely independent of cluster
   radius.  In addition, we have used this
   dataset to search for 
   correlations among the derived parameters.  In particular, we do
   not find any strong trends between the age and mass, mass and effective
   radius, nor between the galactocentric distance and effective
   radius.  There is, however, a strong correlation between the age of
   a cluster and its extinction, with younger clusters being more
   heavily reddened than older clusters. 
   \keywords{Galaxies: individual: M51 -- Galaxies: star clusters}
}
         
\maketitle

%

\section{Introduction}

This study aims at understanding the formation history
of star clusters, their properties and spatial distribution
in the interacting spiral galaxy M51.  To study such a system, one
needs the superb spatial resolution of the {\it Hubble Space Telescope}
({\it HST}) in order to avoid crowding effects and to differentiate between
individual stars, associations, and compact star clusters.

Although much work has already been done on extra-galactic star clusters using
the {\it HST}, most studies have concentrated on specific components
of the full cluster populations (e.g. star clusters in the center of
spiral galaxies, B\"oker et al. 2001) or star cluster systems in
extreme environments such as galactic mergers (e.g. Miller et
al. 1997,  Whitmore et al. 1999).  In order to form a baseline to
study the effects of environment on cluster formation, evolution and
general cluster properties, one must study the properties of clusters
in normal (i.e. more common) environments.  It is only then that we
are able to see which properties are truly unique for a given environment and
which properties remain fixed across all environments.  

One such property of star cluster systems that has a well-established
baseline is that of the luminosity function.  Larsen (2002) has shown
that the cluster populations of spiral galaxies follow a luminosity
distribution that is well represented by a power-law, $N(L) \propto
L^{-\alpha}$, with $\alpha$=2.0.  This is remarkably similar to that
found in all other environments (de Grijs et al. 2003c), and hence can be regarded as a
general property of young star cluster systems.  Recent work by \cite{larsen04}
shows that young clusters in spiral galaxies have typical sizes of
$R_{{\rm eff}} \sim$ 3-10 pc, with the youngest clusters having extended
halos.   It remains to be seen
how other properties, such as mass, age and
spatial distributions
depend on the environment in which the clusters form.

M51 provides an almost ideal astrophysical laboratory to study
extragalactic star clusters.  This is due to its relatively close
distance of $\sim$ 8.4 Mpc (Feldmeier et al. 1997) and its almost
face-on orientation.  Physically it is an interesting case study
because it seems to have had a strong interaction with its
companion, NGC 5195 an S0 peculiar galaxy, during the last few
hundred Myr, which presumably 
caused its grand design spiral appearance as well as its high star
formation rate.  The M51/NGC 5195 encounter(s) have been modelled in
great detail by many authors; this will allow us to compare the
derived age distribution of the star clusters with the orbital
parameters of the system, which are robustly established.  The system
has been modelled relatively 
successfully by a single early passage $\sim$300 $\pm$ 100 Myr ago
(Toomre \& Toomre 1972, Hernquist 1990, and Salo \& Laurikainen
2000, hereafter SL00).  Additionally, SL00 propose a model of a
double passage, the first happening $\sim$400-500 Myr ago and the second,
or last encounter, happening $\sim$50-100 Myr ago.  This latter model
seems to fit many observed details that the single passage models fail
to reproduce, such as the observed counter-rotation of the southern
H{\sc i} tail. 

Much work has already been done on the star cluster system of M51.  A
full literature review is beyond the scope of this paper, but we refer
the reader to the following work:  Bik et al. (2003, hereafter Paper
I) have studied the 
star cluster population in a relatively small area to the north east
of the nucleus that included inner sections of one of the spiral arms.
They find that 
the mass function is reasonably well fit by a power-law of the form
$N(M) \propto M^{-\alpha}$ where $\alpha \sim 2.0$.  They also show
that there is weak evidence for a possible increase in the cluster
formation rate  
in the inner spiral arms at $\sim 400$ Myr ago, 
the proposed time of an interaction between M51
and NGC 5195, but that the evidence is only at the 2 $\sigma$ level.
Boutloukos \&
Lamers (2003) used the data from Paper I to determine the
characteristic disruption timescale of clusters in this region, which
is $\sim$ 40 Myr for a $10^{4} M_{\odot}$ cluster.
Larsen (2000) has 
studied a few of the ``young massive clusters'' which form the high end of the
luminosity function, using combined
{\it HST} and ground-based observations.  He finds that although M51
has a relatively high 
star-formation rate, it fits the specific $U$-band luminosity
($T_{L}^{U}$) vs. star formation rate relation of non-starburst
galaxies.  This has been interpreted as showing that high-mass star
clusters will form whenever the star formation rate is high enough.
Scoville et al. (2001) have used 
many of the same {\it HST} images used in this study, to look at OB star
formation in M51.  They find that the mass function of OB star clusters
is well represented by a power law of the form $N(M_{{\rm cl}})/{\rm
d}M_{{\rm cl}} 
\propto M_{{\rm cl}}^{-2.01}$, in agreement with the results presented
in Paper I.  They also show that although the spiral arms
only make up $\sim 25$\% of the disk surface area they contain a much
higher fraction of the catalogued H{\sc ii} regions.

This study is the second in a series of three, which aim at understanding
the properties of the star cluster 
system of M51 as a whole.  This paper introduces the data and
techniques used to determine the cluster properties, analyses the age
distribution as well as the correlations among the cluster
parameters, namely their age, mass, extinction, size, galactocentric
distance, and position in the galaxy.  In the third paper
(\cite{gieles}, hereafter Paper III) we determine the luminosity and
mass distributions and derive the disruption timescale for clusters
in M51. 

The study is setup in the following way: in \S~\ref{observations} we
present the observations used in this study, including the reduction and
photometry. In \S~\ref{selection} the methods of finding the clusters
and the selections applied to the data are elucidated. In
\S~\ref{fitting} we present our method of 
determining the ages, extinctions, and masses of each of the detected
clusters, and discuss how contaminating sources are
detected and removed. In \S\S~\ref{extinction} and
\ref{agedistribution} the general properties of the cluster 
system are presented and discussed, while \S~\ref{young-rates} presents
evidence for a young, short lived cluster population.  In
\S\S~\ref{sizes} and \ref{correlations} the size distribution and
relations between size and 
the other cluster parameters are presented, followed by
\S~\ref{conclusions} in which we summarize our main results.


\section{Observations, reduction and photometry}
\label{observations}

Our analysis of stellar clusters in M51 is based on observations done
with the {\it Wide Field Planetary Camera-2} (WFPC2) and the {\it
Near-Infrared Camera and Multi-Object Spectrometer} (NICMOS) on board
the {\it HST}. 

\subsection{WFPC2 Observations}
\label{subsec:wfpc2}
Observations in 8 different passbands of the inner region of M51 were
obtained from the {\it HST} data archive: F255W ($\approx$ UV), F336W
($\approx U$), F439W ($\approx B$), F502N ($\approx$ OIII), F555W
($\approx V$), F656N ($\approx$ H$\alpha$), F675W ($\approx R$) and
F814W ($\approx I$).  We emphasize that we have not transformed the
{\it HST} filters into the standard Cousins-Johnson filter system.

\begin{figure}[b]
     \includegraphics[angle=0,width=8cm]{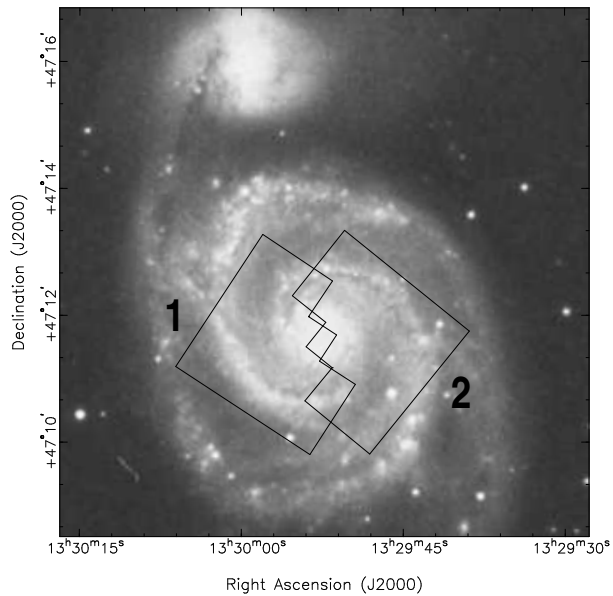}
      \caption{The orientation of the two {\it HST} pointings for the
      F336W, F439W, F555W, F675W, F814W, and F656N filters overlaid on a {\it
      Digital Sky Survey} image of M51.  North is up, and east is to
      the left.  The left
      orientation/pointing is referred to as field 1, while the right
      is referred to as field 2.} 
         \label{fig:orientation1}
\end{figure}

An overview of the filters and exposure times is given in
Table~1. Proposal ID 5777 and 7375 contain the F555W band images and
will be referred to as Orientation 1 and Orientation 2, respectively.
All the other orientations are rotated/shifted to these 2 base 
orientations. An overview of the orientation of the data sets is
shown in Figures~\ref{fig:orientation1} and
~\ref{fig:orientation2}. The combined data sets span a 
region of about 180$^{\prime\prime}$$\times$200$^{\prime\prime}$
(i.e. 7.3 $\times$ 8.1 kpc) containing
the complete inner spiral arms of the galaxy.

The raw images were flat fielded using the automatic standard pipeline
reduction at the Space Telescope Science Institute (STScI). Warm pixels
were removed using the \texttt{warmpix} task which is part of the
IRAF/STSDAS\footnote{The Image Reduction and Analysis Package
(IRAF) is distributed by the National Optical Astronomy Observatories,
which is operated by the Association of Universities for Research in
Astronomy, Inc., under cooperative agreement with the National Science
Foundation. STSDAS, the Space Telescope Science Data Analysis System,
contains tasks complementary to the existing \texttt{IRAF} tasks. In
this study Version 2.1.1 (December 1999) was used.} package. Bad
pixels where fixed with the \texttt{wfixup} task which interpolates
over bad pixels in the $x$-direction.

Cosmic rays where treated differently for the three datasets. For the
images consisting of multiple exposures of the same field per filter
the task \texttt{crrej} was used. This combines exposures of the same
field by rejecting very high counts on an individual pixel basis. Three
iterations were done with rejection levels set to 8, 6 and 4$\sigma$. 

In single exposure images the cosmic rays were removed using the
\texttt{Lacos\_im} package of \cite{2001PASP..113.1420V}. This algorithm
identifies cosmic rays of arbitrary shapes and sizes by the sharpness
of their edges and reliably discriminates between poorly sampled point
sources and cosmic rays. The parameter setting used was a
\texttt{sigclip} of 6.5, a \texttt{sigfrac} of 0.5 and an
\texttt{objlim} of 4. Four iterations were done to give the best
result, because the result of the cosmic ray rejection is very sensitive to
the settings. Careful tests were
done to make sure no real sources were removed by the cosmic ray
removal task. This resulted in almost cosmic ray free images. The few remaining
cosmic rays will be rejected later when the photometry
coordinates from the source detection in the different filters 
will be cross correlated.

\subsection{NICMOS Observations}
\label{subsec:nicmos}

\begin{figure}[b]
     \includegraphics[angle=0,width=8cm]{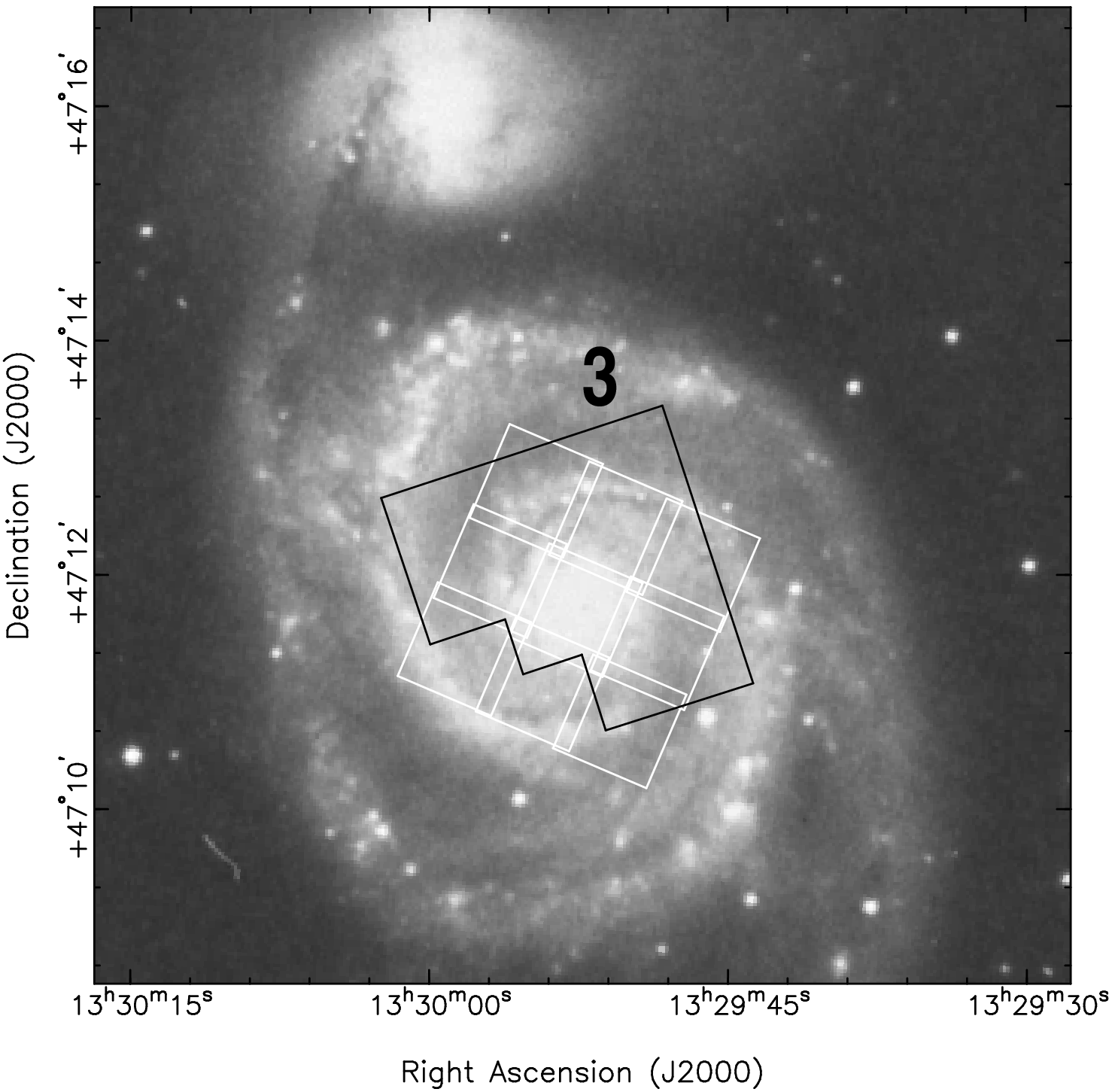}
      \caption{The orientation of the F255W FoV (black) and the NIC3
      mosaic (white).} 
         \label{fig:orientation2}

\end{figure}

NICMOS images were obtained with the NIC3 camera, which has a field of
view (FoV) of about 52$^{\prime\prime}$$\times$52$^{\prime\prime}$. A
mosaic of nine overlapping pointings of the NIC3 camera of M51 was
available in the archive, yielding a total FoV of
186$^{\prime\prime}$$\times$188$^{\prime\prime}$ (i.e. 7.6 $\times$
7.7 kpc) of the central region
in the near-infrared F110W and F160W filters (see
Fig.~\ref{fig:orientation2}). Each of the nine mosaic 
positions was observed using a square dither in each filter setting.

Flat-fielding, dark frame correction and bias subtraction were all
performed in the pipeline image reduction and calibration by the
\texttt{calnica} task. Bad pixels were repaired with the
\texttt{fixpix} task which uses static bad pixel mask images. All
individual images in the dither pattern were shifted to the
same orientation with {\tt imshift}. The dither pattern constituted of shifts of an
integer number of pixels, the images were perfectly aligned on
subpixel level. The individual images in the same orientation were
added with \texttt{imcombine} using the \texttt{crrej} method for
cosmic ray rejection. Finally, the nine co-added images were mosaiced by
matching sources in the overlap region and comparing histograms of the
sky values. This results in a large square FoV overlapping a large
area of the optical data set. Since the mosaic routine
does not scale the count rates values when matching the individual chips,
the mosaic images can be used for source selection and photometry.

\begin{table*}[t]
\caption{Overview of the datasets used.$^{\dagger}$Equivalent to the
V3 Position Angle of the WFPC2.}
\vskip 1mm
\begin{center}
\begin{tabular}{ccccccc}
\hline
ID      & Filters       & Exp. time     &Observation Date       &Orientation angle$^{\dagger}$      \\ \hline
7375    & F336W         & 2 x 600 s.    &21 July 1999           &275.9               \\
(WFPC2) & F439W         & 500 + 600 s.  &                       &                       \\
        & F555W         & 2 x 600 s.    &                       &                       \\
        & F656N         & 1300 + 700 s. &                       &                       \\
        & F675W         & 500 s.        &                       &                       \\
        & F814W         & 700 + 300 s.  &                       &                       \\
\hline
5777    & F439W         & 2 x 700 s.    &15 Jan 1995            &101.5               \\
(WFPC2) & F555W         & 600 s.        &                       &                       \\
        & F675W         & 600 s.        &                       &                       \\
        & F814W         & 600 s.        &                       &                       \\                                             
\hline
5123    & F502N         & 400 + 1400 s. &24 January 1995        &93.2                \\
(WFPC2) & F656N         & 500 + 1200 s. &                       &                       \\
\hline
5652    & F255W         & 4 x 500 s.    &12 May 1994            &333.0               \\
(WFPC2) & F336W         & 2 x 400 s.    &                       &                       \\
\hline
7237    & F110W         & 128 s.        &28 June 1998           &-113.2               \\
(NICMOS3)& F160W        & 128 s.        &                       &                       \\
\hline
\end{tabular}
\end{center}
\end{table*}

\subsection{Source selection}
\label{subsec:sourceselection}

Point-like sources were identified in the individual WFPC2 chip images
and the NIC3 mosaic images with the \texttt{daofind} task from the DAOPHOT
package Stetson (1987). For the optical filters a
threshold of 5$\sigma$ was used, where $\sigma$ is the sky noise. In order not to be biased towards
certain areas on the chips, the value of $\sigma$ was determined on a
low background region on each of the chips.  For the F255W filter the threshold was 4$\sigma$ and
for the near-infrared NIC3 images the
threshold was set to 8$\sigma$. The threshold was chosen such that in each
filter and chip the resulting source density was more or less the
same. No limitation was set on sharpness or roundness, to include as
many sources as possible, including unrelaxed young clusters.

\subsection{Photometry}
\label{subsec:photometry}

\subsubsection{WFPC2}
The coordinates from the source lists, obtained from the
\texttt{daofind} output, were used as an initial guess of the centers
for the \texttt{phot} 
aperture photometry routine.  For the WF chips we used an aperture,
annulus (inner boundary of background annulus) and dannulus (width of
the background annulus) of 3, 7 and 3 pixels respectively.  For the PC
chip we used slightly larger values, namely 4, 10, 3 for the aperture,
annulus and dannulus respectively.  The chosen values for the apertures are
$\sim 3$ times 
larger than the FWHM of the PSF in order to account for the fact that
at the distance of M51, clusters appear slightly extended (i.e. are
not point sources).  Photometric
calibration was done by applying zero-point offsets from Table~28.2 of
\cite{voit} in the VEGAMAG system. 

Total magnitudes were determined
by correcting for flux outside the measurement aperture.  Ideally
aperture corrections are determined from real sources on the
image.  In the case of our data there were not enough isolated bright sources
located in regions of low background to obtain
a meaningful correction.  Thus aperture corrections from the
measurement aperture radius to a 
0.5$^{\prime\prime}$ radius were determined from analytically
generated clusters by convolving a PSF (generated using {\it Tiny
Tim}, Krist \& Hook 1997) with King profiles using the {\it Baolab}
package (Larsen 1999).  King profiles of concentration factor, $c$, of 30 ($c =
r_{t}/r_{c}$ where $r_{t}$ is the tidal radius and $r_{c}$ is the core
radius), and
effective radius, $R_{{\rm eff}}$, of 3 pc (see \S~\ref{sizes}) were
used for the analytic cluster profiles. The values for the three WF
chips are averages. For  
the PC and WF chip in each filter the values are listed in 
Table~3. A final $-0.1$ mag. correction was applied to all sources to correct for
the light missed in the 0.5$^{\prime\prime}$ aperture.  After applying
the aperture corrections, the magnitudes where corrected for CTE loss
according to the equations of \cite{1999PASP..111.1559W}.

\subsubsection{NICMOS}
The photometry procedure for the NIC3 images is essentially the same
as for the WFPC2 images. An aperture, annulus and dannulus of 2, 5 and
3 pixels were used, respectively. The flux was converted to magnitudes in the
VEGAMAG system using 
the PHOTFNU values from the headers. Aperture corrections from the
measurement aperture radius to a 1.0$^{\prime\prime}$ radius were
determined using the same technique as for the WFPC2 images. Values
are listed in Table~2.  A final correction to a nominal infinite
aperture correction of $-0.08$ mag (i.e., multiplying the flux with
1.075) was applied.

\subsection{Incompleteness due to detection limits}
\label{detection-limits}

\begin{figure}
\begin{center} 
  \includegraphics[width=9.0cm]{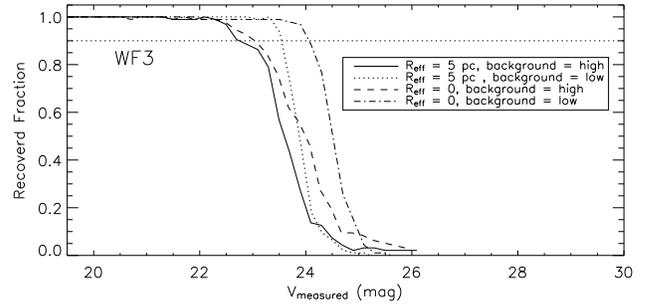}
\end{center}
     \caption{Recovered fraction of sources added to the F555W science
         images on the WF3 chip.  Four examples are shown,
         combinations of low and high 
         background and PSFs and extended clusters ($R_{{\rm eff}} =$ 5 pc).
         }
         \label{example-completeness}
   \end{figure}

\begin{figure}
\begin{center}   \includegraphics[width=9.0cm]{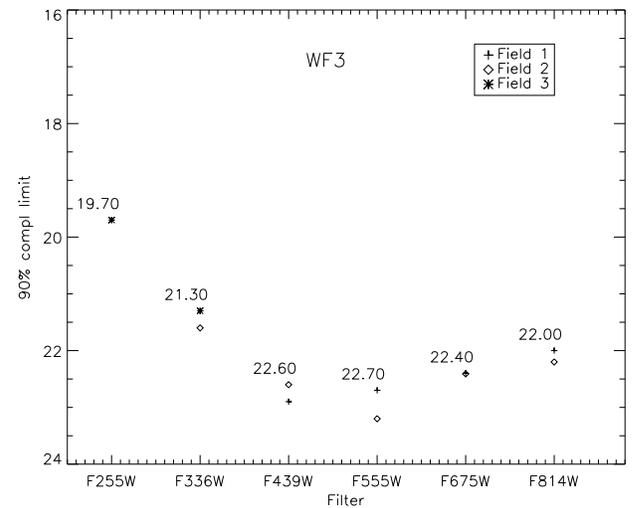}
\end{center}
     \caption{Worst case (high background, $R_{\rm eff}=5$
 pc clusters) 90\%
         completeness limits for PC1 and WF3 for all filters and the
 three available 
         {\it HST} pointings.}
         \label{completeness}
   \end{figure}
        
In order to quantify at what magnitude our sample starts to be
incomplete due to detection limitations, the 90\% completeness limit
was determined for each band for each pointing. To this end, artificial
sources were added to 
the {\it HST} images and were recovered with exactly the same routines as used
to find the real sources. A very important criterion is that objects
should be 5 times above the standard deviation of the background. The
value for $\sigma$ should be the same as when real sources are found. The
magnitude of the added sources were determined and an error criterion of  
$\Delta$mag $<$ 0.2
was applied (see \S~\ref{selection}), the same criterion as
for the real sources to 
enter the photometry list. The tests were done for the PC1 and WF3
chips of both pointings separately.

 Extended sources with different
$R_{{\rm eff}}$ are simulated by convolving WFPC2 PSFs,
produced by {\it Tiny Tim}, with King
profiles. The King profiles are produced by the {\it Baolab}
 package (Larsen 1999). King profiles with a concentration index, $c$, of  30
with different $R_{{\rm eff}}$ were tried. In \S~\ref{sizes} it is
shown that the 
average size for {\it resolved} clusters in M51 is $\sim 3$ pc.  Thus,
we take a more conservative estimate of the completeness survey 
by assuming $R_{\rm eff}=5$
 pc (the more extended the cluster is the more diffuse it becomes, 
for a given brightness,
making it more difficult to detect).  On each chip, two regions were
studied, one with low and one with high background levels.
Since the aperture corrections were derived for average extended sources
($R_{{\rm eff}}$ = 3 pc), the magnitudes of the artificial 5 pc sources are
too faint and the magnitudes of the pure PSFs are
too bright. In both cases the maximum offset is about 0.15 mag.

The recovered fraction as a function of magnitude in the
F555W filter is plotted in Fig.~\ref{example-completeness} for the WF3
chip. Each plot shows the effects of background and source size. The
plots show that for an individual chip the 90\% completeness
 limit varies by more than
a magnitude, depending on the location on the chip and the size of the
source. We therefore do not have a clear detection limit but a region where
incompleteness sets in. Details of this effect and its
influence on the luminosity function are given in Paper III.  The worst case
90\% completeness limits are shown in Fig.~\ref{completeness} for all
bands and for the 3 available {\it HST} pointings. When more than one pointing
is available for a filter, the brightest (i.e. most conservative) 90\%
completeness limit of that filter is chosen.

For the remainder of this study the worst case 90\% completeness limit
was applied to the data. Since the PC chip has been shown to have
many single bright stars (Lamers et al. 2002) and since its
completeness  limit is about 1 magnitude 
brighter than for the WF3 chip, we will only use the WF data for the
remainder of this study.

\begin{table}
\caption{90\% completeness limits for the  WF3 in the 3 different
        available pointings. Most conservative values are printed bold.}
\label{completeness-table}
\begin{center}
\begin{tabular}{ccccccc}
\hline
       & \multicolumn{3}{c}{PC1}        & \multicolumn{3}{c}{WF3}\\ 
Filter & 1     & 2     & 3              & 1     & 2     & 3     \\ \hline
F255W  & -     & -     & \bf{18.80}     & -     & -     & \bf{19.70}\\
F336W  & -     & 20.8  & \bf{20.45}     & -     & 21.60 & \bf{21.30}    \\      
F439W  & 22.10 & \bf{21.80}& -          & 22.90 & \bf{22.60}    & -     \\
F555W  & \bf{21.80}& 21.90     & -      & \bf{22.70}& 23.20     & -     \\
F675W  & \bf{21.20}& 21.90     & -      & \bf{22.40}& 22.45     & -     \\
F814W  & 21.20 & \bf{21.00}    & -      & \bf{22.20}& 22.20     & -     \\
\hline
\end{tabular}
\end{center}
\end{table}

\section{The selection of the clusters and the photometry master list.}
\label{selection}

The coordinate lists for the individual WFPC2 chips were transformed
to one system using the \texttt{metric} task. This task corrects for
geometric distortion and yields coordinates with respect to one frame
(WF3).  Transformation matrices to convert the mosaic
coordinates of field id \#5652 (F255W and F336W) (see Table 1)
 to Orientation 1 and Orientation 2 coordinates,
were determined by finding sources in common to both frames in the F336W
filter. Since the \#5777
 program does not contain an F336W exposure, the
F439W exposure was used instead. The sources used to find the
transformations were selected to be
isolated and spread over the images by as much as possible to make the
transformation as accurate as possible. The spatial transformation
functions were calculated based on the coordinates of the sources in the
two frames using the
\texttt{geomap} task in IRAF. The accuracy of the transformation
is approximately 0.1-0.2 pixels. The same was done for the
near-infrared filters (F110W
and F160W). For these transformations identical sources were found
in the F110W and F814W 
bands. In this case the accuracy was somewhat lower since the NIC3
pixel size ($\approx$0.2$^{\prime\prime}$) is larger then the WF pixel size
($\approx$0.1$^{\prime\prime}$). The accuracy of the transformation
function is still 0.2-0.3 WF pixels.

The final source list was made by cross correlating sources found in the
various filters.  First, a cross correlation between the F439W, F555W
and F675W was 
done. If a coordinate was found in all 3 filters, allowing a mismatch
of 1.4 pixels, it was defined as a genuine source. Then the other
filters were compared 
with the source list, again allowing a mismatch of 1.4 pixels. The
transformed coordinates of the F255W and F336W frame were allowed to
have a 1.6 pixel mismatch. The F110W and F160W frames were allowed to
have a 1.8 pixel mismatch. This was adopted  because of the uncertainty in the
transformation.

The resulting initial source list contains 3504 objects that are
  observed in at least the F439W, F555W, and F675W bands above the $5
  \sigma$ detection limit. Most of these
  will be eliminated later in the study of their energy distributions
  because the photometry is uncertain or the energy distribution
  does not match that of cluster models (see \S~\ref{elimination}).

\begin{table}[t]
\caption{Aperture correction values for all filters (mag).}
\vskip 1mm
\begin{center}
\begin{tabular}{cccc}
\hline
Filter  &PC                     &WF3 (mean)     &NIC3\\ \hline
F225W   &$-0.32$         &$-0.14$         &               \\
F336W   &$-0.31$         &$-0.13$         &               \\
F439W   &$-0.31$         &$-0.13$         &               \\
F555W   &$-0.32$         &$-0.13$         &               \\
F675W   &$-0.32$         &$-0.13$         &               \\
F814W   &$-0.33$         &$-0.14$         &               \\
\hline
F110W   &                       &                       &$-0.22 \pm 0.01$ \\
F160W   &                       &                       &$-0.22 \pm 0.01$ \\
\hline
\end{tabular}
\end{center}
$^{\dagger}$ This value is applied to our sample.  For clusters with
$R_{\rm eff}=5$ pc, these values decrease by $\sim 0.15$ mag. (e.g. to
-0.28 for F555W in the WF3 chip).

\end{table}

\subsection{Contamination of the sample by stars}
\label{contamination}

  We estimate the possible number of massive stars that may
  contaminate our source sample by scaling the star population of the
  LMC, measured by Massey (2002) to the conditions in M51.
  This sample, which is complete down to $V\simeq 15.7$ mag or $M_V \simeq
  -3.2$ mag covers a total area of 10.9 kpc$^2$ and includes most of
  the bar, the 30 Doradus region and several other well populated fields.
  This area contains more than half of the mass of the LMC.  
  We have calculated the intrinsic visual magnitudes, by adopting
  only foreground extinction of $A_V=0.40$ mag and $E(B-V)=0.13$ mag
  (Massey et al. 1995)
  and a distance modulus of 18.48 mag (Westerlund 1997) 
  The total number of stars brighter than $M_V<-7.0$ mag
  (which corresponds to $V < 22.60$ at the distance of M51, for
  no extinction) is less than 100, and the number with $M_V < -8.0$ mag
  is 6. By far the majority of these stars are found in clusters.
  We can get an estimate of the fraction of the luminous {\it field} stars
  compared to the total fraction from the data in Table 12 of Massey
  (2002) which gives the number of massive stars outside
  clusters. (This is the only large and homogeneous extragalactic
  stellar sample deep enough to determine the relation between the
  populations of field stars and the total populations.)
  He found 186 field stars (i.e. outside known OB associations) 
  more massive than 40 \Msun, which corresponds to $M_{\rm bol}
  \le -9$ mag. More than 90 per cent of these are on or near the
  main sequence, where $T_{\rm eff}>25~000$ K and hence the 
  bolometric correction is so large that $M_V>-7.0$ mag. At most 10\% 
  of these stars, i.e. less than 18, are optically bright at 
  $M_V<-7.0$ mag in the LMC. 
  
  With a total mass of the observed area of about
  $4 \times 10^9~\Msun$ (Meatheringham 1991), this
  implies about 4 to 5 field stars more 
  massive than 40 $M_\odot$ per $10^9$ \Msun. (This number could be
  slightly higher if the dark matter content of the LMC within the region
  used here is considerable). 
  This small number is due to the very steep IMF of the 
  massive field stars. Massey (2002) showed that the stellar IMF of
  massive field stars can be approximated with $\Gamma \simeq -4$,
  compared to $\Gamma \simeq -1.3$ for cluster
  stars, where $\Gamma = 1 + {\rm d}(\log N(M))/{\rm d}
  (\log(M))$.  

The total luminous mass of M51 within a radius of 5 kpc is $2.1 \times
10^{10} M_{\odot}$ (Salo \& Laurikainen 2000).  Scaling the number of
field stars per unit mass in the LMC to that of M51, we estimate that
at most 170 to 210 field stars with $M_V<-7.0$ will be in our field,
and less than
  15 to 20 with $M_V<-8.0$, {\it if there were no extinction in M51}.
  Even a moderate extinction will remove these stars from our sample.

  Another estimate of contaminating stellar sources in our sample
  can be obtained by comparing the number of field stars to that of
  observed clusters.  This comparison relies on the assumption that
  the ratio of field stars to clusters is the same in different
  galaxies.  This assumption will not be valid for low or intermediate
  mass stars between galaxies with widely different cluster disruption
  timescales, as galaxies with shorter disruption timescales will have
  a higher field star to cluster ratio.  However, for young massive
  stars (lifetimes less than 10 Myr), disruption is not expected to
  significantly influence this ratio.  Using the catalog of Bica et
  al. (1996) we see that there are $\sim$ 180 clusters and
  associations brighter than $M_{V} = -6.94$ in the LMC (assuming a distance
  modulus to the LMC of 18.48 and an imposed completeness of $m_{V}$ =
  22.70 at the distance of M51).  Using the arguments given above,
  there are $< 18$ young massive stars that are brighter than $M_{V} =
  -7.0$ in the LMC.  Thus there are less than one field star for every
  ten clusters in the LMC.  Therefore, using this approximation, we
  expect less than 10\% of our M51 sample to be bright young stars.

   A third alternative method to estimate the contamination by massive
isolated field stars is  to scale the number of isolated field stars
in the LMC to that of M51 using the mass normalized star formation
rates.  In the studies summarized by Grimm et al. (2003), M51 has
about half the star formation rate per unit galaxy mass as the LMC.
So, using this assumption we expect half as many bright field stars in
M51 as in the LMC per unit mass, which results (using the mass ratios
above) in $\sim 47$ bright field stars.

  We conclude that the number of contaminating stars in our sample is
  small\footnote{Bik et al. (2003) found that massive young clusters
  in the inner NE-spiral arm of M51 have a surprisingly small flux in
  the O[III] $\lambda 5007$ line.  They interpret this as evidence
  that there is a lack of massive stars of $M > 30 M_{\odot}$ in the
  clusters.  If this is also valid for the field stars, the number of
  contaminating field stars in our sample will be even smaller than
  estimated here.}.  Moreover, we will show in Section
  \ref{elimination} that most of
  these will be removed from the sample because they do not pass the
  criterion that their energy distribution can be fitted with a cluster
  model.

\section{Determination of the cluster parameters from their energy distribution}\label{fitting}

 The age, mass and extinction of star clusters can be derived by
  comparing their spectral energy distributions to those of cluster
  evolution models (e.g. Paper I, de Grijs et al. 
  2003a,b,c). This method can also be used to 
eliminate contaminating sources (e.g. stars and background galaxies).

\subsection{Adopted cluster models}
\label{models}
We have adopted the updated GALEV simple stellar population (SSP) models
(Schulz et al. 2002, Anders \& Fritze v. Alvensleben 2003)
as our templates to derive ages, extinction values, and masses for each of
the observed clusters in M51.  These models use the isochrones from
the Padova group, which seem to fit the data better than the Geneva
tracks (Whitmore \& Zhang 2002).  The set of models that we use
assumes a Salpeter stellar IMF from 0.15 - 50 $M_{\odot}$.  We
have used models with metallicities of 0.4, 1, and 2.5 
$Z_{\odot}$, and a comparison between the different metallicities will
be made in Section~\ref{metallicityeffect}.

There were two major reasons for selecting the updated GALEV tracks.
The first is the inclusion of gaseous emission lines as well as
continuum emission, which can
severely affect the broad-band colours for young star clusters 
(see Anders \& Fritze-v. Alvensleben 2003).  The
second is that they have been published in the {\it HST} WFPC2/NICMOS filter
system, which eliminates possible inaccuracies in converting the {\it HST}
observations to the standard Cousins-Johnson system.

\subsection{The fitting of the spectral energy distributions}
\label{3def}

We used
the three-dimensional maximum likelihood fitting method ({\it 3DEF})
developed
by Bik et al. (2003).  The method has been described in detail there,
and hence we will only give a brief summary of it here.  In passing we
note that this method has been tested against colour-colour methods of
age fitting, and has shown itself to be superior (e.g., de Grijs et
al. 2003a vs. Parmentier, de Grijs \& Gilmore 2003).

The method works as follows: we use a grid of
SSP models, i.e. the GALEV models (see previous section) that give the
broad-band colours 
of a single age stellar population as a function of age.  For each
model age, we apply 50 different extinctions in steps of 0.02 in
E($B-V$).  We adopted the Galactic extinction law of Savage and
Mathis (1979), which was found to agree with the extinction law derived
for M51 by Lamers et al. (2002).  We 
then compare the model grid with the observed spectral energy
distribution of the clusters using a minimum $\chi^{2}$
test, with each observation weighted with its uncertainty.  The model (age
and extinction) with the lowest $\chi^{2}$ is selected, and the range
(i.e. maximum and minimum age) of accepted values is calculated by
taking the most extreme model that satisfies $\chi^{2}_{\nu} <
\chi^{2}_{\nu,{\rm min}} + 1$.  Once the age and extinction have been
calculated, the mass is found by comparing each observed filter
magnitude to each model magnitude (iteratively with respect to age and
extinction).  Since we have first scaled the GALEV models to a {\it
present mass} of $10^{6} M_{\odot}$  the difference
between the observed and the predicted magnitude (taking into account
the distance modulus and extinction) can be converted to the present
mass of the observed cluster\footnote{All masses derived in this paper
refer to the {\it present mass} of the clusters, not to their {\it
initial mass.}  If stellar evolution is the only mass loss mechanism of
the cluster, the initial mass of the clusters can be retrieved by
correcting the present mass for the fraction lost by stellar
evolution, given by the GALEV models.}.

\subsection{Test of the accuracy of the method}
\label{sec:accuracy}

We have tested the accuracy of the method used for the determination
of the age and mass of clusters. To this purpose we created a grid of
synthetic clusters, distributed evenly in logarithmic age, in the
range of $7.0<\log (t/{\rm yr})<10.0$ in 31 steps, and evenly in logarithmic
mass in the range of $3.0 < \log (M/{\rm M}_\odot)<6.0$ in 21 steps. This results in
651 synthetic clusters.
The clusters have a distance modulus of
$m-M=29.64$, as for M51, and were assigned a random extinction
of $0 < {\rm E}(B-V) < 1.0$ mag with a Gaussian probability function that peaks at
E$(B-V)=0$ mag and has a $\sigma=0.20$ mag. For each of these
clusters we calculated the expected energy distribution in the
same {\sl HST/WFPC2} filters as used for this study of M51, using the
GALEV cluster models with a metallicity of $Z=0.02$, and a Salpeter
IMF. The  photometry of each synthetic cluster was given  
an uncertainty that depends on magnitude. The uncertainty was derived
from the observations, which show that $\Delta(m)$ can be approximated
quite well by $\Delta(m) = 10^{d_1+d_2 \times m}$, where $m$ is the magnitude
and the values of $d_1$ and $d_2$ are empirically determined 
for each {\it HST} filter. These values are listed in Table
\ref{deltamag}. We then added a random correction to the magnitude
of each cluster in each filter, within the interval  ($-\Delta(m)$,
$+\Delta(m)$). This resulted in a list of 651 synthetic clusters with  
photometry and photometric errors similar to those of the observed
clusters in M51.

\begin{table}[t]
\caption[]{Empirical parameters for the uncertainty of the magnitudes:
$\Delta(m)=10^{d_1+d_2\times m}$}
\label{deltamag}
\vskip 1mm
\begin{center}
\begin{tabular}{ccc}
\hline
Filter & $d_1$ & $d_2$ \\
\hline 
F255W & $-8.540$ & +0.390 \\
F336W & $-7.850$ & +0.325 \\
F439W & $-8.320$ & +0.328 \\
F555W & $-7.800$ & +0.300 \\
F675W & $-7.600$ & +0.300 \\
F814W & $-8.230$ & +0.328 \\
\hline\
\end{tabular}
\end{center}
\end{table}

\begin{figure}
\begin{center}   
 \includegraphics[width=8cm]{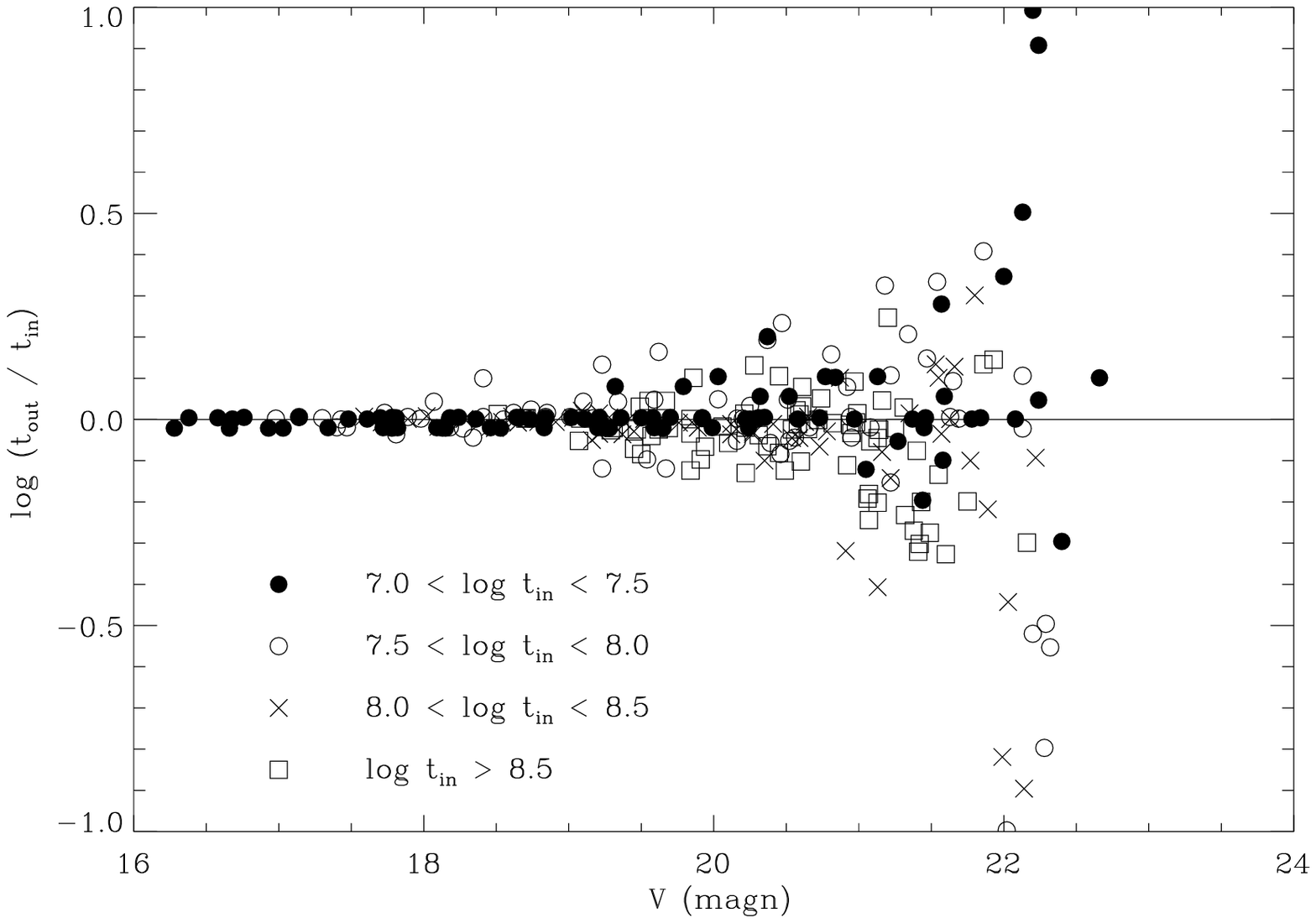}
 \includegraphics[width=8cm]{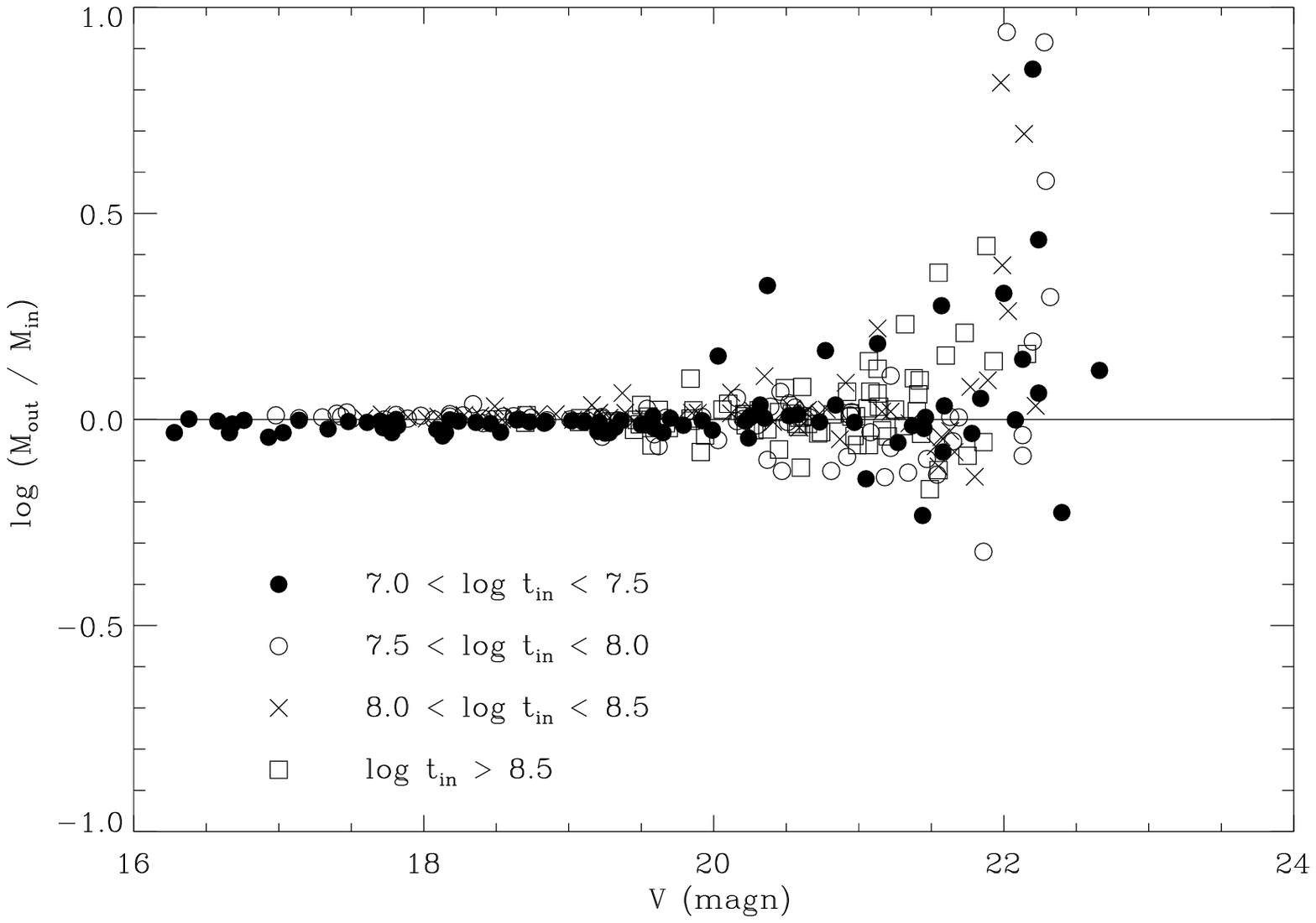}
 \includegraphics[width=8cm]{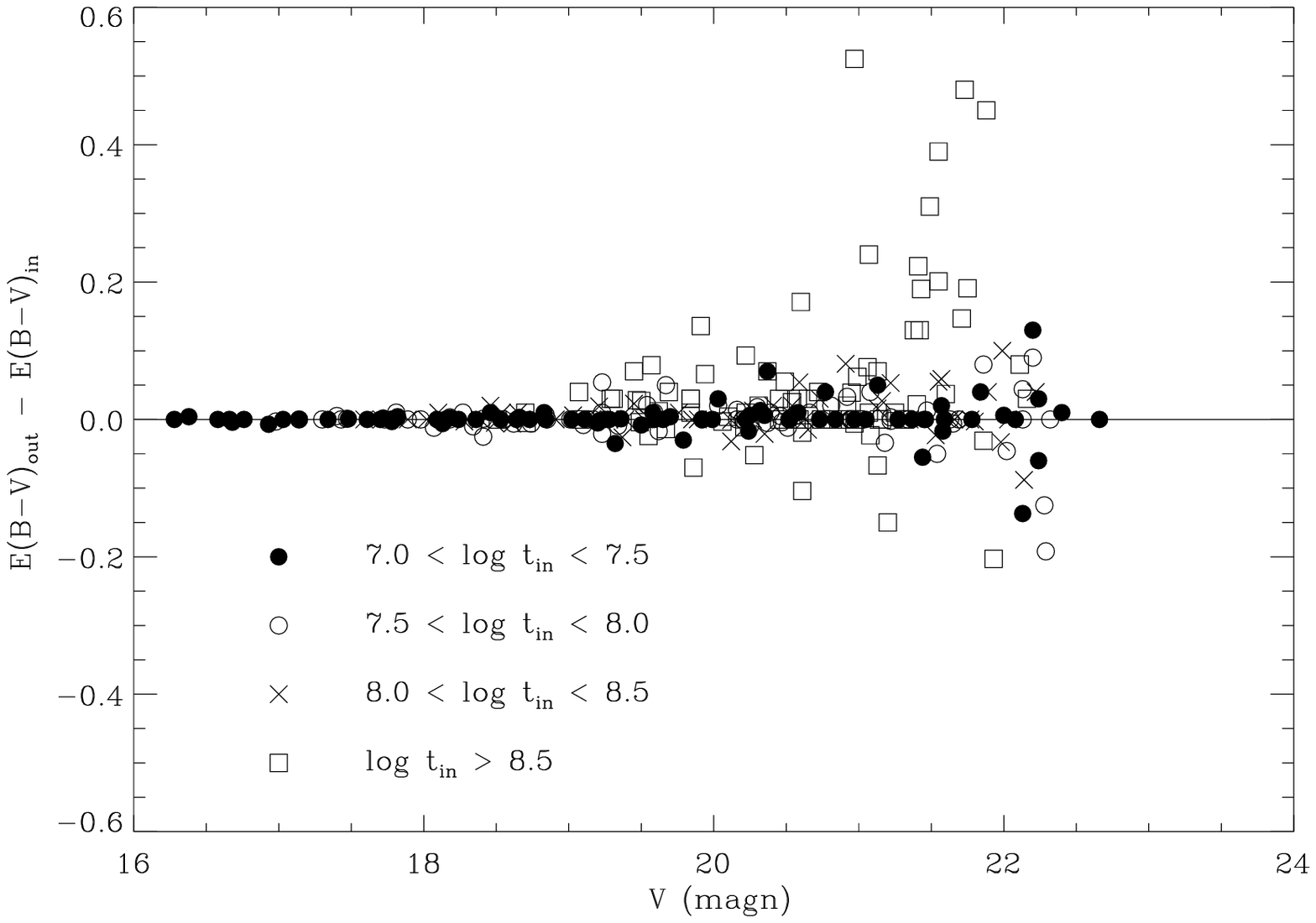}
\end{center}
\caption[]{Comparison between input and output values of the ages
(upper panel, a), masses (middle panel, b) and extinction values (lower
panel, c)
of the synthetic cluster sample. Different symbols indicate 
different age ranges.}
\label{fittest}
\end{figure}

We then ran these synthetic clusters through our cluster fitting
program to determine the age, mass and extinction in exactly the same
way as done for the observed clusters. We adopted the same conservative
magnitude limits as for the observed sample, as given in Table
\ref{completeness-table}. The comparison between the input mass, age and
extinction of the synthetic clusters and the result of the fitting 
gives an indication of the accuracy and reliability of the results.
The result is shown in Figure \ref{fittest}.

The panels show the difference between input and output values of
$\log(t)$, $\log(M)$ and E$(B-V)$ as a function of the $V$ magnitude of the
synthetic cluster. This parameter was adopted because the 
accuracy of the fitting is expected to depend on the accuracy of the 
input photometry, for which $V$ is a good indicator. We have 
divided the results into four age bins (with parameters indicated in
the figure), because the accuracy of the fit may be different in
different age intervals, as the spectral energy distribution depends
on age.

Figure \ref{fittest}a shows that the age determination is quite
accurate, to within about 0.20 dex, for relatively bright 
clusters with $V < 21.0$ mag. For fainter
clusters with $V>21.0$ mag the uncertainty increases up to a maximum of
about 1 dex for clusters with $V>22.0$ mag. The scatter is approximately
symmetric around the zero-line with two exceptions. \\
(i) For the youngest faint clusters with $t<$ 30 Myr and $V>21.5$ mag
the age is more likely to be 
overestimated than underestimated. This is the consequence of the simple
fact that the two youngest GALEV cluster models have ages of 4 and 8
Myr. So a cluster cannot be fitted to a model younger than 4 Myr,
which limits the possible negative values of 
$\log (t_{\rm out}/t_{\rm in})$. \\
(ii) Faint clusters with ages of
$\log(t/{\rm yr})>7.5$ have a higher probability to be fitted with a younger
cluster model than with an older cluster model. This might reduce the
number of old clusters derived from the fitting of the SEDs of
observed clusters. This effect will be taken into account in the analysis
of the observed mass and age distributions of the M51 clusters.

Figure \ref{fittest}b shows that the mass determination is quite
accurate for relatively bright clusters with $V < 21.0$ mag. For fainter
clusters with $V>21.0$ mag the uncertainty decreases up to a maximum of
about $-0.5$ dex to +1.0 dex for the faintest clusters. 
The scatter is approximately
symmetric around the zero-line, except for the tendency to
overestimate the mass of the faint old clusters. This is a direct
result of the underestimate of the age.

Figure \ref{fittest}c shows that the extinction can be derived quite
accurately for young clusters with $\log(t/{\rm yr})<8.5$ and $V<21.5$ mag. For
fainter clusters there is a symmetric scatter up to about
$\Delta {\rm E}(B-V) \simeq 0.1$ mag. For clusters older than 300 Myr there is a
tendency to overestimate the extinction. This is the reason for the 
overestimate of the mass and the underestimate of the ages of these
old clusters.

In general the ages and masses derived from fitting the cluster models
are quite accurate: more than
99\%  of our synthetic clusters have an age determination that is
accurate to within 0.25 dex and a mass determination that is accurate to
within 0.2 dex. For real cluster samples the accuracy will depend on
the magnitude distribution of the clusters, and hence on their
photometric accuracy.

\subsubsection{Stochastic colour fluctuations}
\label{fluctuations}
An additional uncertainty in the fitting of ages of star clusters
using SSP models, is the stochastic colour fluctuations resulting from
the fact that young low mass clusters will be dominated by a few high
luminosity stars (e.g. Dolphin \& Kennicutt 2002).  Thus, the absence
(or presence) of a small number
of stars can significantly affect the observed integrated colours of a
cluster.  Unfortunately, due to the random nature of this effect it is
impossible to determine which clusters in our sample are most heavily
affected by this.  However, if the stochastic colour fluctuations vary
equally around the 'true' (i.e. fully sampled, high mass limit)
colour, we do not expect our derived age distribution for the
population as a whole to be affected by this.  Finally, we note that
our age distribution analysis (see \S~\ref{agedistribution}) has been
restricted to clusters with derived masses above $10^{4.7}
M_{\odot}$, e.g. the most massive clusters, where the stochastic
colour fluctuations are expected to be smallest (Lan{\c c}on \&
Mouhcine 2000).  Stochastic effects are expected to affect the colour
and magnitude of a 10 Myr, $10^{5} M_{\odot}$ cluster by $\sigma_{V}$ = 0.08
mag and $\sigma_{V-I}$ = 0.07 (Dolphin \& Kennicutt 2002).  These
small fluctuations are within 
the observational uncertainties, and as such are not expected to
significantly affect the derived parameters.

\subsubsection{Dependence on the adopted models}
\label{adoptedmodels}
The {\it 3DEF} method assumes that the adopted models accurately
reproduce the colours of star clusters as a function of age.  In order
to quantify this effect we have carried out the same analysis as
discussed in \S~\ref{sec:accuracy}, but now using the
Starburst99 models (Leitherer et al. 1999) for the artificial
clusters, and then carrying out our fittting procedure using the
GALEV models.  These models were chosen because of the differing
stellar evolutionary isochrones used when constructing their
respective SSPs (i.e. the GALEV models adopt the Padova isochrones,
while the Starburst99 models adopt the Geneva isochrones).  

These tests resulted in 80\% of the artificial clusters being fit
within 0.5 dex of the input age and mass.  The extinction is fit to
within 0.2 mag in E(B-V) for $>$ 80\% of the artificial clusters.
Thus we see that while there are significant differences between the
models used, the parameters of the majority of clusters can be
accurately recovered.

We also note that fitting our cluster sample using the Starburst99 models
as templates, resulted in age and mass distributions which are
consistent with the results presented here (Bastian \& Lamers 2003).
Namely, the high numbers of very young ($<10$ Myr) as well as the
increase the cluster formation rate at $\sim 50-100$ Myr ago (see
\S~\ref{agedistribution}).

\subsection{Reduction of the sample based on energy distributions}
\label{criteria}
Once we have obtained the best fit for each source (i.e. age,
extinction and mass) we can then
determine whether or not the object has a {\it good} fit, i.e. if it
is a cluster.   In previous studies (Paper I, de Grijs et al. 2003a)
we used the selection 
criterion that the reduced $\chi^{2}$ must be below 3.0.  This was
effective in eliminating most of the poor fits, but it also removed
many of the brightest clusters because their photometric error was
relatively low.  In order to remove this problem, we have used the
standard deviation, defined to be
\begin{equation}
\sigma^2(BVR) = \sum_{i}\frac{(m^{{\rm mod}}_{i} - m^{{\rm
obs}}_{i})^{2}}{n_{{\rm filters}}}
\label{sigmacrit}
\end{equation}
where the sum runs over the F439W, F555W, and F675W bands only (the
three bands in which all clusters were detected), as a criterion
to test the accuracy of the photometric fit.  The sum was
restricted in order to maintain a fairly homogeneous set and to
eliminate biases that would occur if bands with short
exposures or low count rates (e.g. F255W) were included.

We then checked by eye the spectral energy distributions of the
sources and their fits, in order to find a reasonable value for the
minimum $\sigma^2(BVR)$, $\sigma^2_{{\rm min}}(BVR)$.  As the
``goodness'' of the 
fit is rather arbitrary, we quantified the exact value of $\sigma^2_{\rm
min}(BVR)$ by
making a histogram of all the sources, approximating the histogram
with a Gaussian (which is a relatively good fit) and taking the 1/e
point.  This point, and the value preferred by our visual inspection
were almost identical at $\sigma^2_{\rm min}(BVR) = 0.05$.

In the remainder of the
paper, when we refer to ``clusters'', we will mean sources that passed
the following criteria:\\
\begin{itemize}
\item detected in the F439W, F555W, and F675W bands at at least
5$\sigma$ above the background,  
\item photometric uncertainty $<$ 0.2 mag in the F439W, F555W, and F675W bands,
\item detected in at least one other band,
\item well fit ($\sigma^2(BVR) < 0.05$) by an SSP model,
\item brighter than the 90\% completeness limits for each
filter in which they are detected (see Table~\ref{completeness-table})
\end{itemize}

1150 clusters pass these criteria, which will be used to study the
cluster population in M51.  The majority of the sources (from the
original 3504 sources detected) that have been removed from our sample
were eliminated because they were fainter than one or more of our 90\%
completeness limits. 

\subsection{The elimination of stars on the basis of their energy distribution}
\label{elimination}

In \S~\ref{contamination} we argued that the expected number of very
luminous stars 
that might contaminate our sample is relatively small.  We have investigated
the rejection of possible stellar sources  
from our sample on the basis of the energy distribution fitting. 
To this purpose we created
energy distributions of stellar atmosphere models and analysed them in the
same way in which we analysed the energy distributions of clusters.

We adopt the supergiant models calculated by
Bessell et al. (1998), transformed to WFPC2 magnitudes by Romaniello
et al. (2002). Bessell et al. (1998) normalized the magnitudes to
stars 
with a radius of 1 $\Rsun$ at a distance of 10 pc. They published 70 models
with $T_{\rm eff}$ ranging from 3500 to 50000 K for a range of
gravities from main sequence stars to supergiants. We adopted the energy
distributions 
of these stellar models with the lowest gravity (without adding noise to their photometry) and
then analysed the data as if the stars were 
clusters. We did this for stars that are observed in all six {\it WFPC2}
bands (UV,$U,B,V,R,I$), and for stars observed only in the $U,B,V,R$ bands or in the
$B,V,R,I$ bands. 

We found that stars with
effective temperatures higher than about 14,000 K all end up in the youngest
age bin for clusters of $\log(t/yr) \simeq 6.6$.  If our sample is
contaminated by massive 
stars, most of these will be blue, because they spend most of their
time on or near the main sequence. If there is a red phase for such
massive stars, the crossing time between the blue
and the red phase is very short, only about 1\% of the
main-sequence phase. However stars with masses in excess of 50 $\Msun$
are not expected to become red supergiants because that would bring
them above the observed Humphreys-Davidson luminosity upper limit (Humphreys \&
Davidson, 1979). 
We conclude that, {\it if} there is a sizable number of
massive O-stars with $21.70<V<22.70$ mag, they will end up in the youngest
age bin of our cluster sample.

\begin{figure}
\begin{center}   
 \includegraphics[width=12cm]{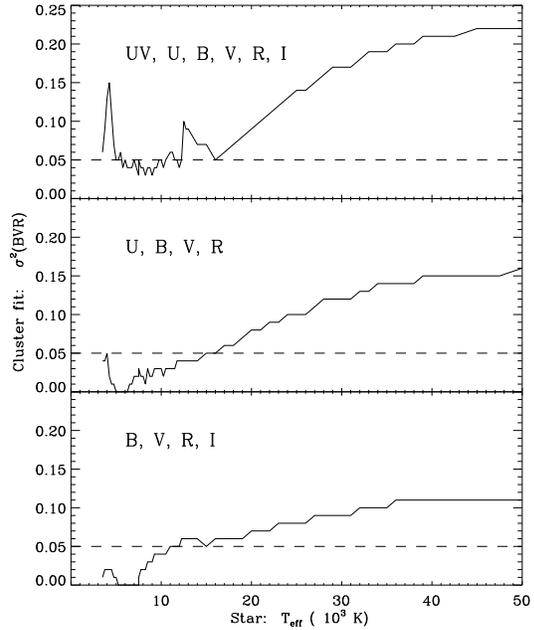}
\end{center}
\caption[]{The elimination of possibly contaminating bright stars 
in our sample by means of the fitting of their energy distribution.
The figure shows the values of the rejection criterion $\sigma^2(BVR)$ (Eq.~1) 
for stars of different temperatures in three different filter
combinations. 
Notice that possibly contaminating stars with $T_{\rm eff} > 17~000$ K will be
effectively eliminated by our selection criterion $\sigma^2(BVR)<0.05$.
}
\label{distinction}
\end{figure}

Fig. \ref{distinction}
shows the values of $\sigma^2(BVR)$ as a function
of the temperature of the stars, for stars observed in all six {\it WFPC2} bands, and for
stars observed only in the $B,V,R,I$ or $U,B,V,R$ bands.
Only stars below the dashed line pass our criterion of
$\sigma^2(BVR)<0.05$.  
We see that most of the stars will be rejected from our sample, except
the coolest ones.

We concluded before (Sect. \ref{contamination}) that the expected 
contamination of
our cluster sample of sources with $V<21.70$ mag with bright stars is very small.
Here we have shown that our analysis of their energy distribution will
eliminate a major fraction of these. The small number that may pass our
selection criterion, are expected to end up in the youngest age bin.

\subsection{The effect of metallicity}
\label{metallicityeffect}

We have studied the effect of metallicity 
on the determination of the age, mass, and extinction of the
clusters. To this purpose we follow the method of de Grijs et al. (2003a,b) and
introduce metallicity into our models as a free parameter.

  This technique has been applied recently 
to NGC 3310 (de Grijs et al. 2003b) with good results, and to NGC 6745
(de Grijs et al. 2003c) with somewhat less encouraging results as two
metallicities were equally well represented in the final sample,
due to the smaller number and narrower wavelength coverage of
available passbands, and the inherent effects of the age--metallicity
degeneracy.  For 
NGC 3310 the authors report that the near-infrared colours of a cluster
are a good indicator of metallicity, while optical colours do a better job at
discerning ages (see also Anders et al. 2004).  

To this end we have fit all sources in M51 which pass our criteria (see
\S~\ref{criteria}) with models of four metallicities, namely  Z =
0.004, 0.008, 0.02, and 0.05.  We do not 
find any significant difference in the number of sources that pass our
selection criteria ($\sigma^2(BVR) < 0.05$) between the different
metallicity models.  We find that 14\%, 21\%, 17\%, and 48\% of the
sources in our sample are best fit (lowest $\sigma(BVR)^{2}$) by
models of Z = 0.004, 0.008, 0.02, and 0.05 respectively.  The largest
percentage of the clusters being fit by twice solar metallicity models
is consistent with the metallicity estimates of H{\sc II} regions in
M51 by Hill et al. (1997).  

We note however, that many of the individual clusters within the same
cluster associations show widely different metallcities.  As this
seems unphysical, we are forced to conclude that fitting on
metallicity is not possible with the current data.  We therefore
choose to concentrate our analysis on the fits using the solar metallicity 
models (which is the median metallcity found in our sample), although
we will discuss how our results depend on the assumed metallicity where
applicable. 

\subsection{Recovery rates}
\label{accuracy}
   In order to determine the recovery rate of the fitting method we
   use the artificial cluster sample generated in
   \S~\ref{sec:accuracy}.  In Fig.~\ref{fittest} we found that the accuracy
   of the fit depended on the magnitude of the input cluster, with
   brighter clusters being fit with higher accuracy.  Because of the
   input cluster IMF, which has many more fainter clusters than
   brighter ones, we need to quantify this affect.
   The recovery rate is defined to be the number of clusters found
   divided by the number of input clusters for a given age bin.
   Figure~\ref{recovery-rates} shows the recovery rate 
   as a function of age.  This shows that we 
 slightly over-predict the number of clusters in the age range 
   of $\log (t/{\rm yr}) \in$ [6.62,6.87] and $\log (t/{\rm yr})
   \in$ [7.37,7.62] by about 40 per cent.  We 
   underestimate the number of clusters in the age range of 
   $\log (t/{\rm yr}) \in$ [6.87,7.12] by about 20\%, and also
   tend to underestimate the number of clusters with ages greater than
   $10^8$ yr. 
   This is not a physical effect but due
   to the rapidly changing colours of the SSP GALEV models during this
   age range. We will use this recovery rate in \S~\ref{cfh} where we
   study the cluster formation history of M51.

{\bf
\begin{figure}
\begin{center}   \includegraphics[width=8cm]{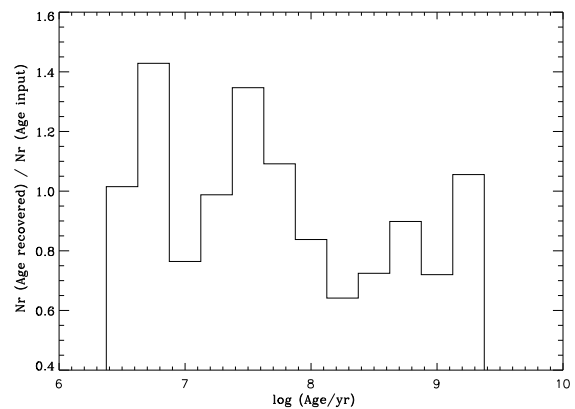}
\end{center}
     \caption{The recovery rate of our sample as a function of input
   age. We underestimate the number of clusters in the range of
   $6.9<\log (t/{\rm yr}) < 7.1$ by about 25\% and overestimate the number
   in the preceding age bin by $\sim 40$\%.}
         \label{recovery-rates}
   \end{figure}
}

        \section{Extinction}
        \label{extinction}

        Once we have derived ages, extinctions, and masses for all of
        the clusters, we can begin looking at general trends in the
        cluster population.  

\begin{figure}
\begin{center}   \includegraphics[width=9cm]{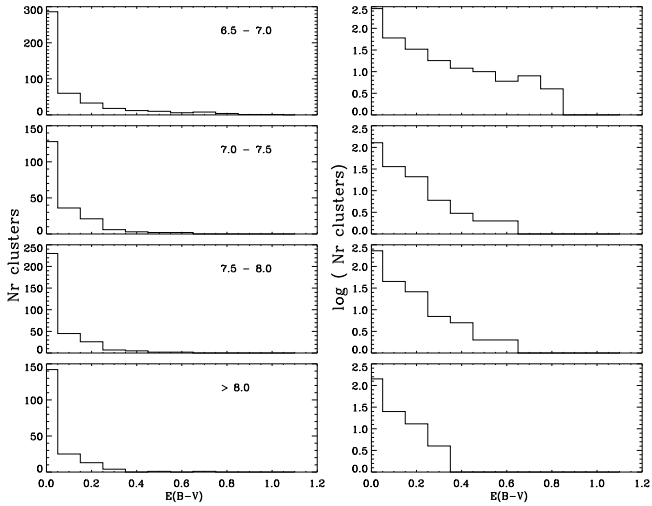}
\end{center}
     \caption{ The histograms of the extinction for different age
        bins (given in the upper right of the panels in the left
        column, in logarithmic years). The left-hand panels show the linear
        distribution ($N$
        versus E$(B-V)$) and the right-hand panels show the logarithmic
        distribution ($\log(N)$ versus E$(B-V)$). Notice the strong peak
        near E$(B-V)\simeq 0$ mag and the steeply declining relation 
        towards E$(B-V) \simeq 0.8$ mag.}
         \label{extplots}
   \end{figure}

        Figure \ref{extplots} shows the distribution of the E$(B-V)$
        values for different age groups. The left-hand side of the figure
        shows the number of clusters and the right-hand side shows their
        logarithmic values. The figure shows that at each age bin
        over  half of the detected clusters have an extinction of
        E$(B-V)<0.05$ mag, with the rest showing a rapidly declining
        distribution up to E$(B-V) \simeq 0.8$ mag. 
        The logarithmic figures show about the same distribution
        of the clusters with E$(B-V)>0.05$ mag, with
        $d \log(N)/d{\rm E}(B-V) \simeq -2.0$.  This relation is
        independent of age, although the highest extinction found
        decreases strongly with ages.  This is presumably due to the fading of
        clusters as they age, bringing them closer to the detection
        limit, so older clusters with a significant amount of
        extinction will drop below the detection limit.  Another
        effect that could contribute to the observed distribution is if
        young clusters have more extinction than older clusters, which
        is expected as young clusters form in gas rich environments.
   
        The large numbers of clusters found with low extinction is
        qualitatively what
        is expected. The
        cluster IMF of $N(M) \sim M^{-\alpha}$, with $\alpha \simeq
        2$, implies an increase in cluster numbers towards the lower
	mass and fainter clusters. The magnitudes of these low mass
        clusters are close to the detection limit, so they can only   
	be detected if their extinction is small.  Thus our sample is
        complete to larger extinction values for brighter clusters,
        and for younger ages which are less affected by the detection
        limit (see Fig.~\ref{age-vs-mass}).  This effect will affect
        all cluster studies where the host galaxy is not optically
        thin (i.e. studies where one cannot see through the host
        galaxy at the given completeness limits).

\begin{figure}
\begin{center}   \includegraphics[width=9cm]{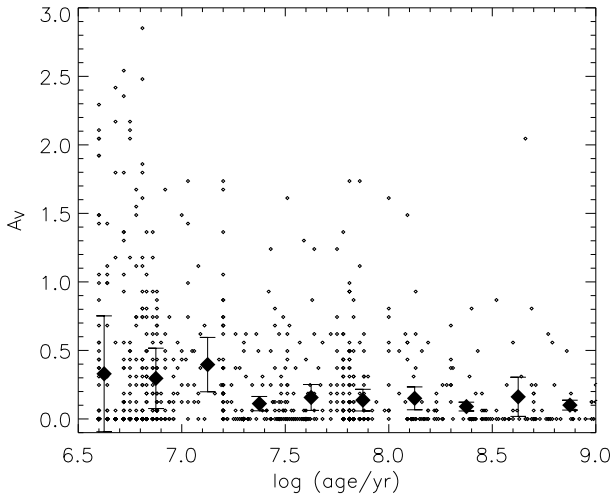}
\end{center}
     \caption{The extinction as a function of age for the clusters in
         our sample.  The solid points are the mean for each age bin, while the
         error bars are the variance about the mean.  Note that the
         extinction is higher for clusters with ages less than 20
         Myr than for the older clusters.}
         \label{age-vs-e}
   \end{figure}

	In Fig.~\ref{age-vs-e} we show the extinction as a function of
	age for the clusters in our sample.  The solid points are the
	mean of all the points for each age bin and the error bars
	represent the variance about the mean.  This figure shows that
	the average extinction is higher for young clusters ($\leq
	20$ Myr) and rather abruptly drops to a constant value for
	clusters older than 20 Myr.  This is qualitatively what is
	expected as clusters emerge from their parent molecular cloud,
	although as we noted above, there is a selection effect which
	causes us to detect younger clusters to higher extinction values.

   \section{The age distribution of the clusters}
   \label{agedistribution}

        It has been shown that the global properties of the host galaxy,
        such as an interaction or merger, can significantly affect the
        cluster formation rate (CFR) within a galaxy (e.g. de Grijs et
        al. 2003a).  To search for such an effect within M51, we
        combined our cluster sample, with the detailed models of the
        interaction between M51 and its companion, NGC 5195, taken from
        the literature (SL00).
        SL00 propose a scenario in which NGC
        5195 has had two close passages with M51, an early encounter
        at approximately 400-500
        and a second at 50-100 Myr ago (the last passage).  Their {\it
        double encounter} model 
        reproduces the kinematic features of the system better than
        the single passage models (such as, e.g., the apparent counter-rotation
        of the southern HI-tail), as well as the morphological ``kink''
        in the northern spiral arm.

        \subsection{The mass versus age diagram}
        \label{mtdiagram}

 \begin{figure}
   \includegraphics[width=9.0cm]{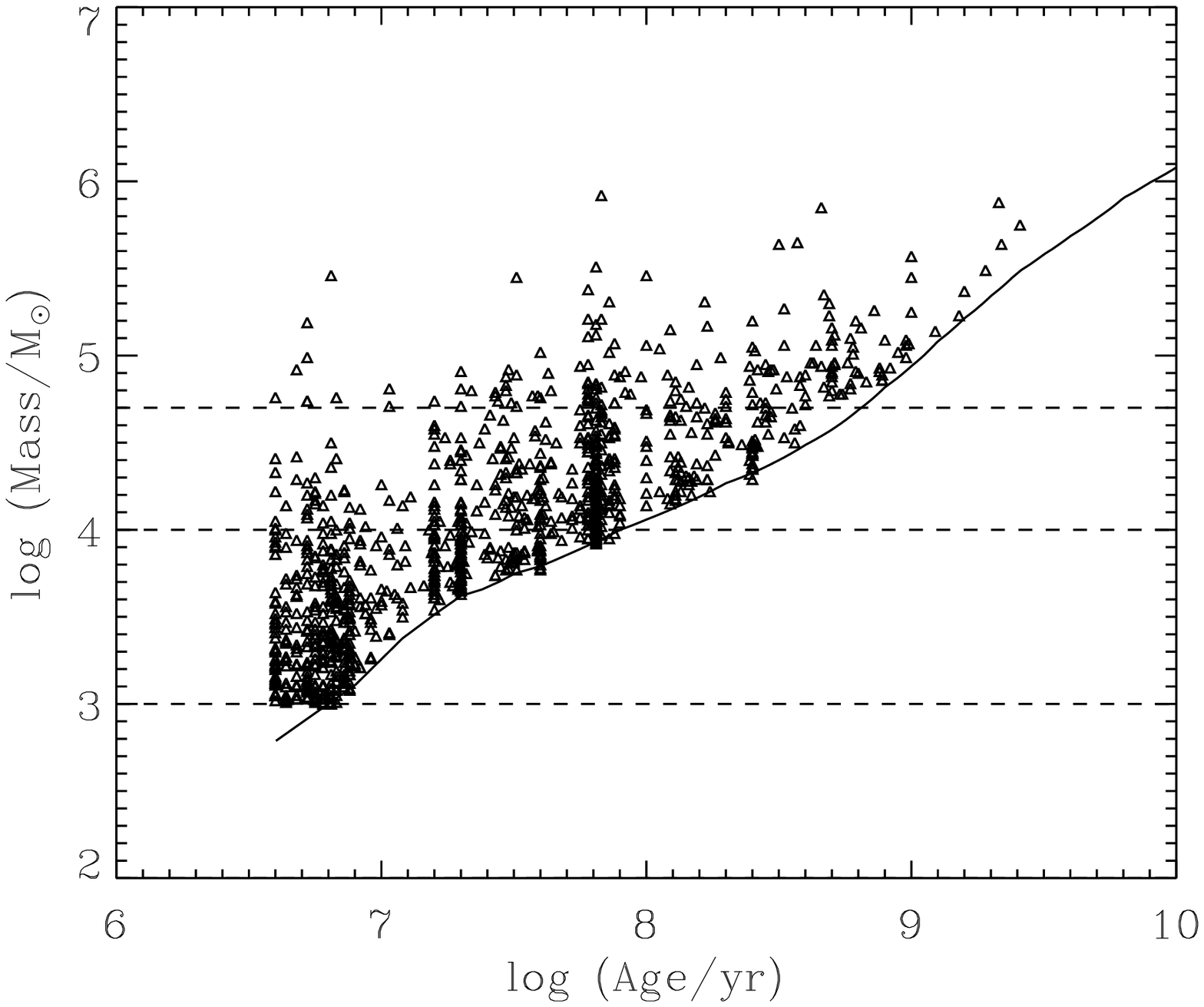}
   \includegraphics[width=9.0cm]{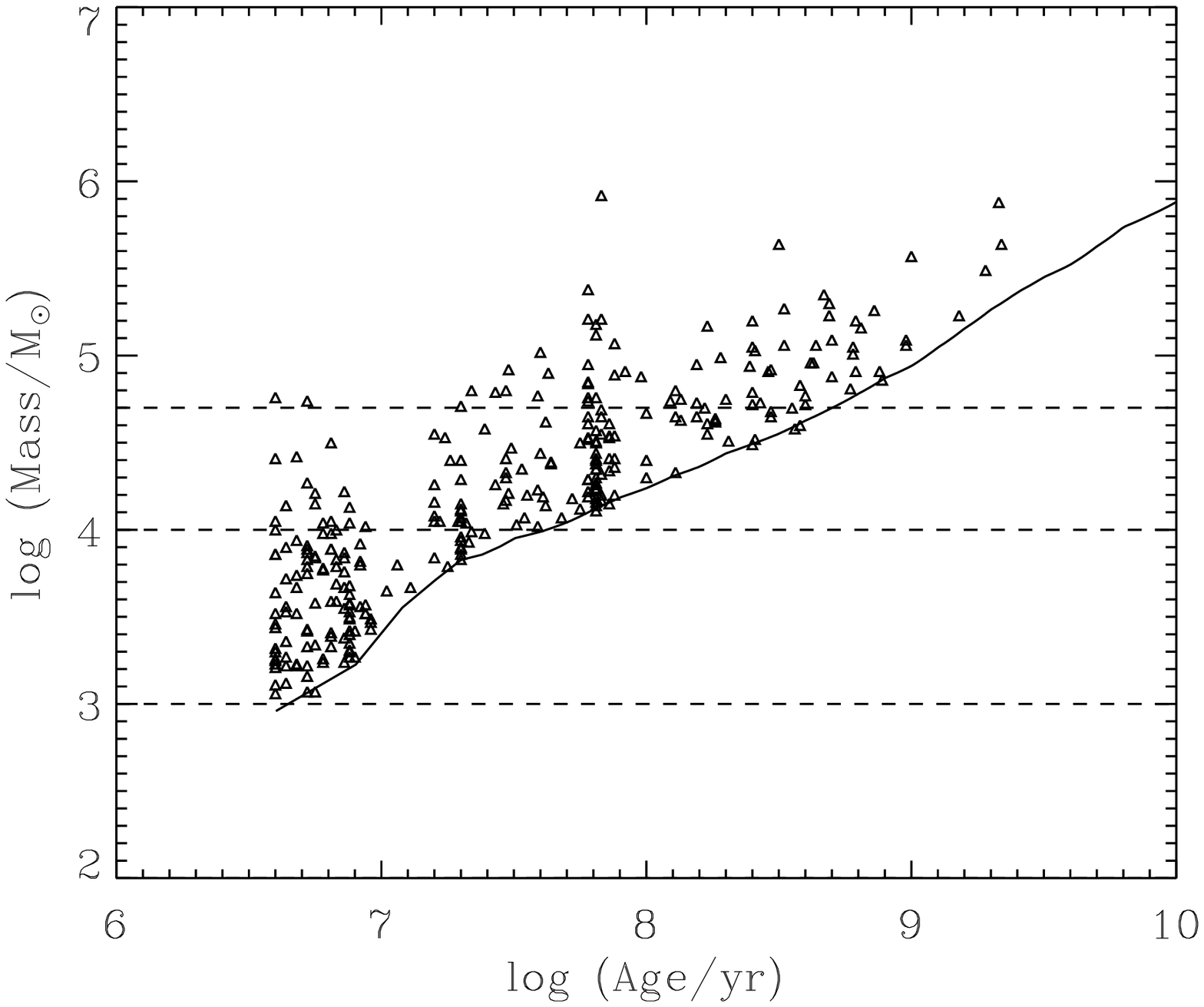}
   \includegraphics[width=9.0cm]{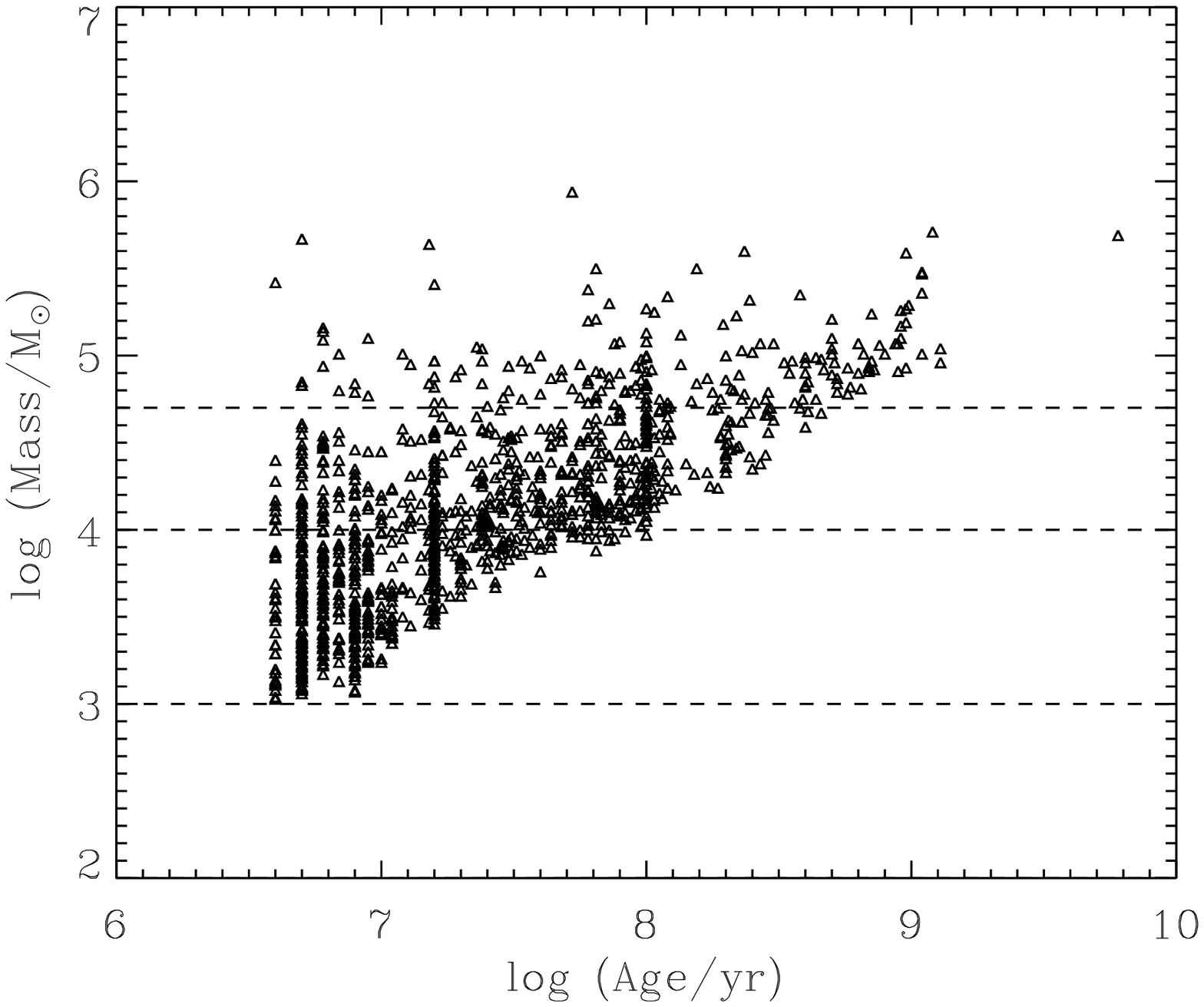}
      \caption{{\bf Top:} The present mass as a function of age for
   the 1078 clusters 
   that pass our $\sigma$ criterion and which have a present mass
   greater than 1000 
   $M_{\odot}$.  The solid line is the
   predicted detection limits using the GALEV models, for m(F439W) $=
   22.6$ mag.  The horizontal dashed lines 
   represent constant mass cutoffs used in Fig.~\ref{age1}. {\bf
   Middle:} The same as above, but only for those clusters with
   measured sizes greater than 2 pc.  The solid line is the predicted
   detection limit for m(F555W) $= 22.0$ mag {\bf Bottom:} The present mass as
   a function of age while leaving metallicity run as a free parameter.}
         \label{age-vs-mass}
   \end{figure}

        Figure \ref{age-vs-mass}, top panel, shows the age of each cluster
        plotted against its determined mass. (We remind the reader
        that this is the present mass of the cluster.)  
         Note the presence of a clear lower mass limit which
   increases with age. This is due to the evolutionary fading of clusters
   as they age, which raises the minimum mass that the clusters should
   have to be brighter than our detection limit.  
   Hence, the lack of old low-mass
        clusters is mainly an effect of fading due to stellar
        evolution (another important effect is cluster disruption,
        which will be 
        treated in Paper III).
        The solid lines increasing to 
        the right show the detection limit for $m_{\rm F555W}$ = 22.7 mag,
        i.e. the adopted magnitude limit of the sample, and 21.7 mag.

        The figure shows the apparent  
        concentrations of the clusters in some very narrow age bins,
        i.e. at $\log (t/yr)$=7.2, 7.3 and near 7.8. This is not a
        physical effect but it is due to the way we select the best
        fitting models in the study of the cluster SEDs with the 3DEF-method
        (see \S~\ref{sec:accuracy}  and Paper I).  In reality
        these peaks should spread over the neighbouring age bins.

	The bottom panel in Fig.~\ref{age-vs-mass} shows the relation
	between the age and masses of those clusters whose sizes are
	resolved.  The general properties are the same as for the full
	sample, namely the increase in the lower mass limit as a
	function of increasing age, an excess of clusters with young
	ages ($<$ 10 Myr), and an additional excess at 60 Myr.
	Therefore we conclude that we are not biasing our study by
	including unresolved sources.  

	As mentioned above, certain artifacts of the age vs. mass
	distribution are likely caused by model and fitting effects,
	which do not reflect anything physical.  Since this complicates
	the analysis, we attempt to correct for these effects using
	the simulations from \S~\ref{accuracy}.  To do this, we
	generate a number density grid in the age vs. mass space for
	our observed cluster sample as well as for our simulated
	cluster sample (cluster IMF with slope $-2.0$, continuous cluster
	formation rate, and random extinction between $0 <$ E$(B-V) <
	1$ mag with a
	Gaussian probability function that has $\sigma=0.20$
	mag).   The model number density is normalized in order that
	the average is the same as the observations (i.e. we normalize
	to the cluster formation rate).  We
	then divide the observed cluster distribution by that of the
	simulated distribution.  This technique will remove any
	features caused by the adopted models or the fitting method.
	
\begin{figure}
   \includegraphics[width=9cm]{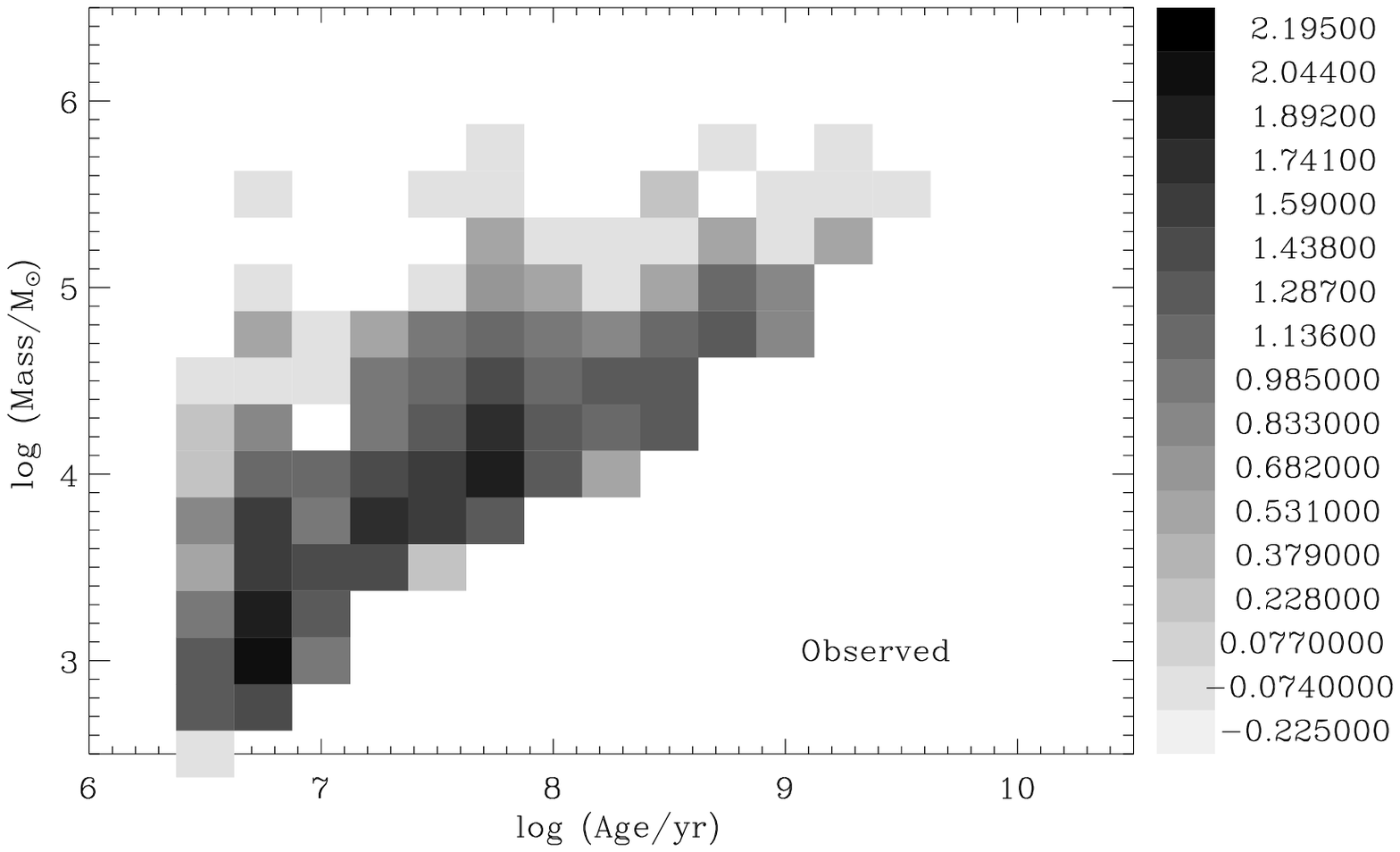}
   \includegraphics[width=9cm]{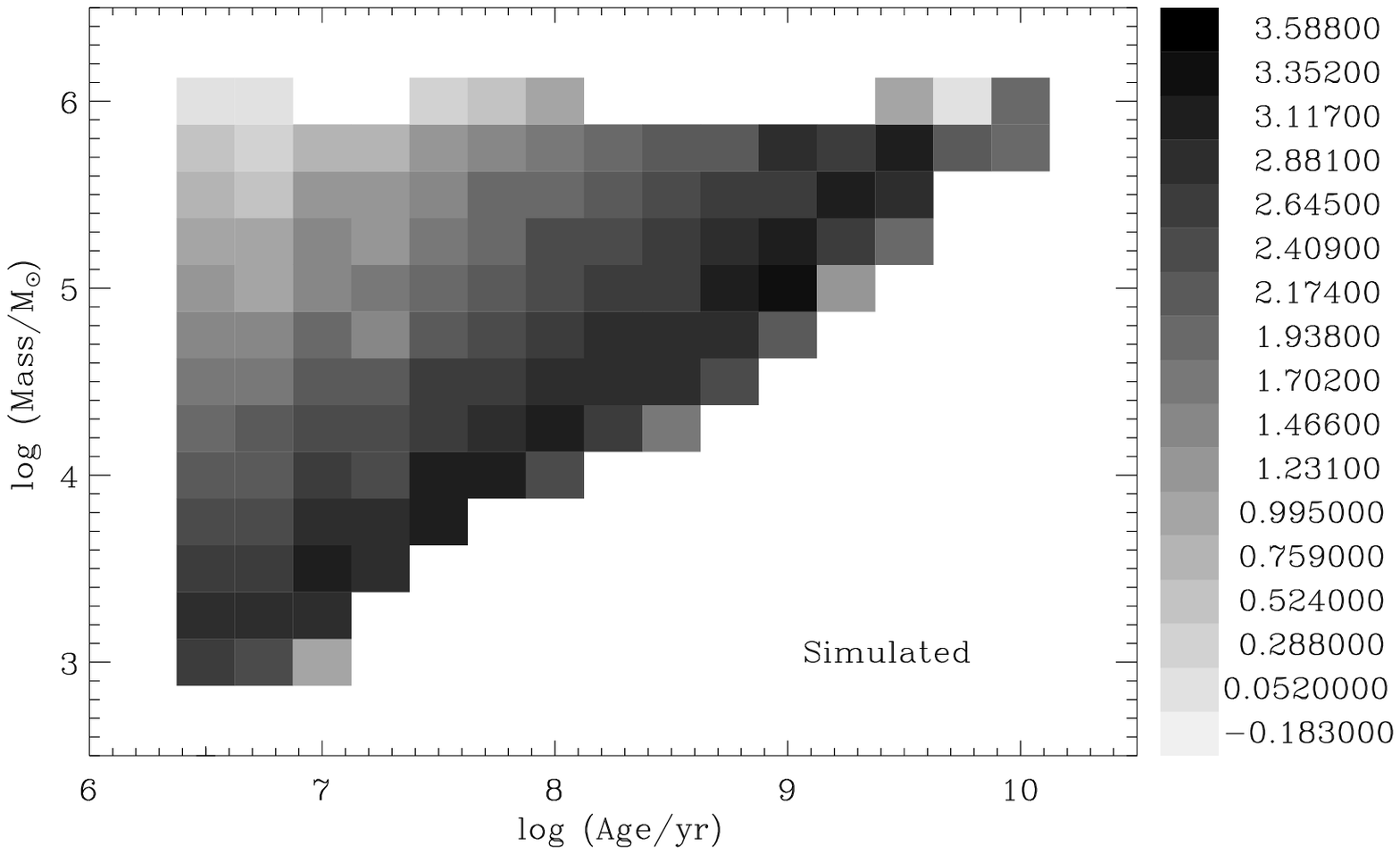}
   \includegraphics[width=9cm]{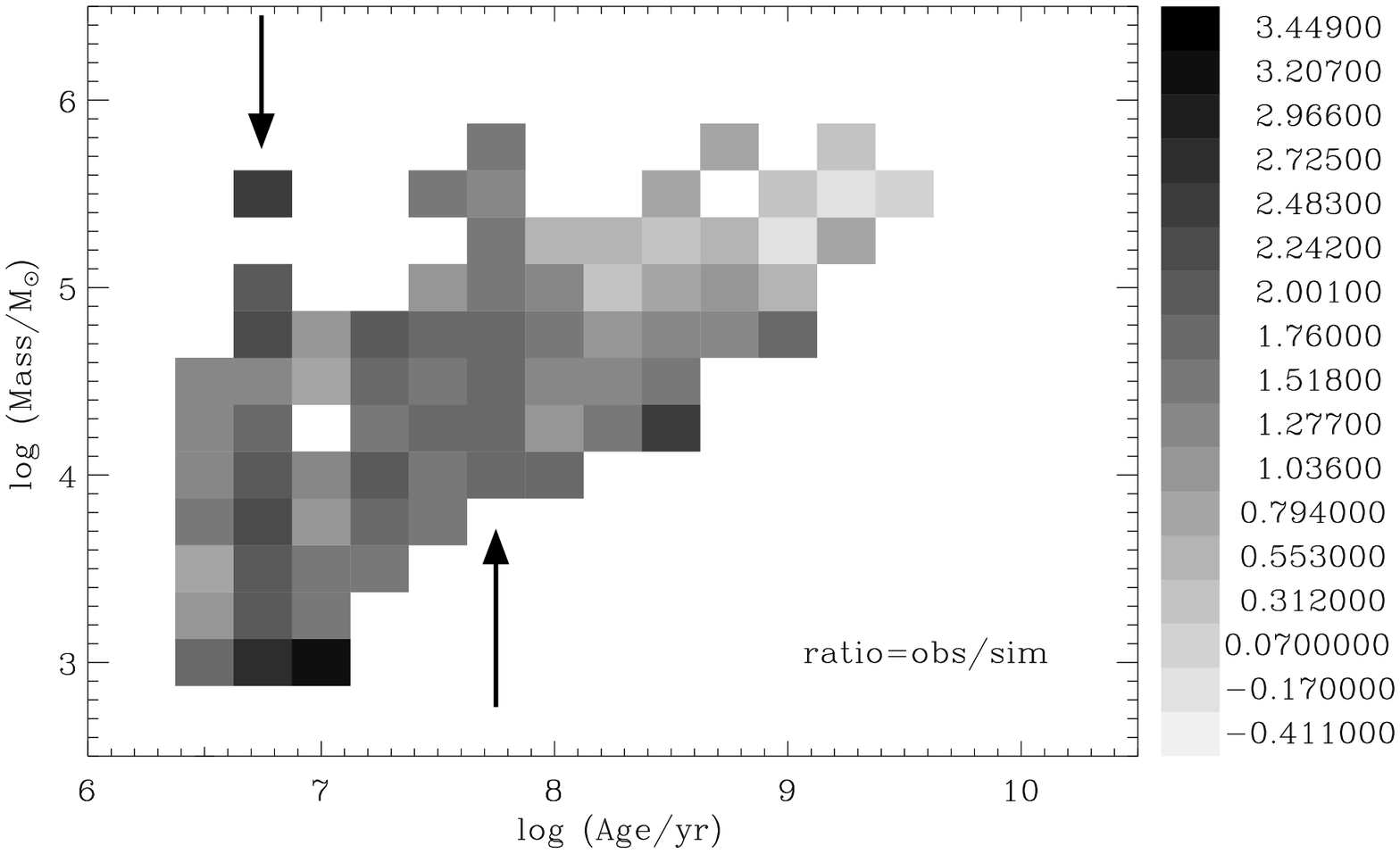}
      \caption{{\bf Top(a):} The present mass as a function of age for
   the 1152 clusters 
   that pass our selection criterion.  The shading represents the
   logarithmic number
   density, with the scale given on the right.  {\bf
   Middle(b):}  Same as (a), but for the simulated cluster sample with
   a  constant cluster formation rate (see \S~\ref{sec:accuracy} for
   details). {\bf Bottom(c):}  The ratio between the observed
   distribution (top) and the predicted distribution (middle).  The
   arrows indicate two age bins which have an overdensity of clusters
   relative to the simulated cluster sample.} 
         \label{age-vs-mass-grey}
   \end{figure}

	This process and the result are shown in
	Fig.~\ref{age-vs-mass-grey}.  It shows the number density of the
	clusters in the mass versus age diagram in a logarithmic
	grey-scale, normalized in such a way that the number densities
	run from light to dark grey.  The top panel shows the density
	of the {\it observed} clusters in the mass versus age
	diagram.  The middle panel shows the density of the {\it
	generated} cluster sample with a constant cluster formation
	rate and a power law cluster IMF.  The lower panel shows the
	{\it ratio} of the observed to predicted cluster number
	densities in the diagram.  For the top and middle panels,
	dark shading represent high numbers of clusters, while for the
	bottom dark shading represents an over-density of observed
	clusters compared to that expected from the models.  In order to
	fully interpret the final result, it is important to
	understand the simulated cluster sample.  Two of the strongest
	features in Fig.~\ref{age-vs-mass-grey}b are the increase in
	the number density from the top of the panel to the bottom as
	well as from left to right.  The increase in the number of
	clusters at lower masses (top to bottom in
	Fig.~\ref{age-vs-mass-grey}b) is due to the cluster initial
	mass function, which gives many more low-mass than high-mass clusters.
	The increase in the number (for a given
	mass bin) of clusters from younger to older (left to right) is
	due to the logarithmic age-binning and the assumed constant
	cluster formation rate in linear time.  

	Once we divide our observed distribution
	(Fig.~\ref{age-vs-mass-grey}a) by the simulated distribution
	(Fig.~\ref{age-vs-mass-grey}b) we have the {\it true} age vs
	mass distribution which is shown in Fig.~\ref{age-vs-mass-grey}c.
	Since the effects of the cluster IMF and the logarithmic
	binning have been removed in this comparison, the decrease in
	number density from
	the lower left of Fig.~\ref{age-vs-mass-grey}c to the upper
	right is presumably caused by the disruption of the clusters.
	Additionally we find relative excesses in the number density at ages
	of $\sim 60$ Myr and $\sim 6$ Myr.  These two features will be
	discussed in more detail in \S~\ref{cfh} and
	\S~\ref{young-rates}, respectively.

 \subsection{The cluster formation history}
 \label{cfh}

\begin{figure*}
\begin{center}   \includegraphics[width=15cm]{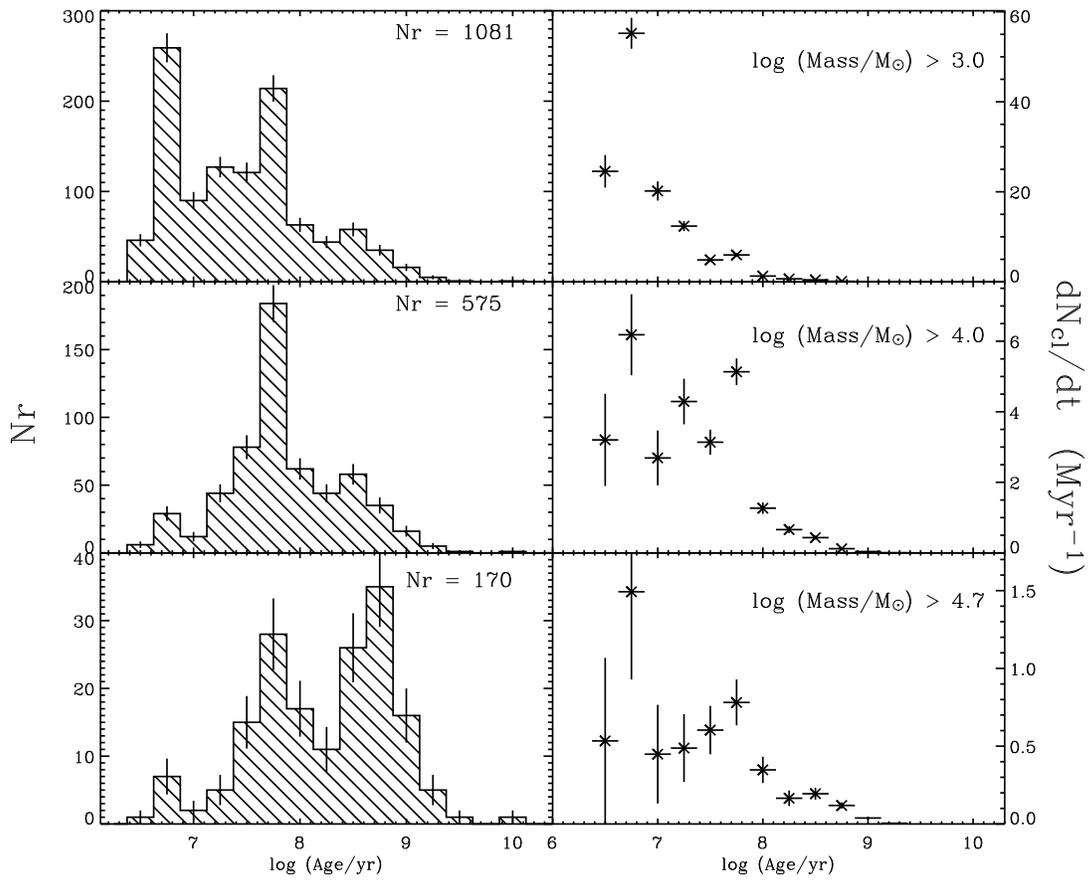}
\end{center}
     \caption{Differences between the age distribution and the
         formation distribution.  Left: the number of clusters found
         per age bin for three mass limits.  Right: the cluster
         formation rate (number per Myr) as a function of age,
         corrected for the recovery errors given in
         Fig.~\ref{recovery-rates}. 
         Adopting different mass limits clearly biases the interpretation of
         the distribution.}
         \label{age1}
   \end{figure*}

	In the previous section we found an excess of clusters at
	ages $\sim$ 60 Myr, corresponding in time to the {\it last}
	passage of NGC 5195 and M51 in the SL00 models.  In order to
	quantify this excess we will look at the age distribution
	directly (i.e. by collapsing the mass axis in
	Fig.~\ref{age-vs-mass-grey}c). As shown in \S~\ref{mtdiagram},
	for a given magnitude-limited sample,  we are able to detect
	clusters to lower masses for
	younger ages than for older ages.   Therefore, we will
	detect more younger than older clusters (assuming the
	standard cluster IMF which predicts many more lower mass
	than higher mass clusters), simply due
	to selection effects.  Thus, in order to
	trace the true age distribution of the sample, the analysis
	must be restricted to masses above a certain limit, where the
	sample is complete for the age range of interest.  This effect
	is shown in Fig.~\ref{age1} by looking at the age distribution
	and applying the constant mass cuts shown in
	Fig.~\ref{age-vs-mass}.  The mass cuts are at 
	10$^{3}$, 10$^{4}$, and 5$\times 10^{4} M_{\odot}$, the last one
	chosen so that we are complete for ages younger than 1 Gyr.

	The left-hand side of Fig.~\ref{age1} shows the number of clusters
	detected per logarithmic age bin while the 
	CFR (in number per Myr) is shown on the right-hand side.  For
	the calculation of the CFR, we have corrected the age
	distribution of the detected clusters for the recovery rates
	(Fig.~\ref{recovery-rates}) in order to eliminate spurious
	effects.  The  
	difference in appearance between the two sides is due to the
	logarithmic binning of the left-hand side (e.g. each bin spans a
	larger age range than the preceding bin).  Note that the
	distributions for different mass cuts appear very different.

        If we look at the sample in which we are complete for the past
        Gyr (i.e. for masses $> 10^{4.7}$ M$_{\odot}$) we see a clear
        indication of an increase in the CFR $\sim$ 60 Myr
        ago.  In addition, we also find the peak at 
        $\sim 6$ Myr, that was seen in
        Fig.~\ref{age-vs-mass-grey}c. This peak is presumably due to the
        presence of a large 
        number of unbound clusters, and will be discussed in
        detail in \S~\ref{young-rates}. 

        The increase in the CFR at $\sim 60$ Myr ago for clusters with 
        $\log (M/M_\odot) > 4.7$  
        occurs at the time of the proposed second encounter between
        M51 and NGC 5195.  The CFR appears to have been 2 to 3 times higher
        during the burst period than in the period immediately preceding
        it.  The CFR declines slightly after the initial rise, but
        remains significantly higher than before.  In Fig.~\ref{age1}
        we note the CFR rises extremely rapidly at the onset of the
        burst, and decays much more smoothly, consistent with a linear
        Gaussian distribution plotted on a logarithmic scale.  
        
	\subsubsection{Duration of the burst}

         An increase in the CFR at the time of an 
        interaction has been shown for other galaxies, e.g. M82,
        NGC 3310 and NGC 6745 (de Grijs et al. 2003a,b,c,
        respectively).  While the duration of the burst of cluster
        formation in M51 is rather short lived, we note that this is similar
        to the distribution observed in NGC 3310 (de Grijs et
        al. 2003b) where the estimated duration of the increase in
        cluster formation rate was $\sim 20$ Myr (it should
        be noted that their method
        of analysis was very similar to the one presented here).
        Presumably, the increase in the CFR in 
        M51 was likely caused by the tidal interaction between M51 and
        NGC 5195 $\sim 60$ Myr ago, supporting the {\it two encounter}
        model of SL00.  

	Tidal interactions can influence large portions of a galaxy
        within a relatively short time, and is
        the only known mechanism that could coordinate
        galaxy-wide starbursts on the timescales presented here.
        Recently, Murray et al.(2004) have shown that there is an
        Eddington-like 
maximum luminosity for starburst galaxies. This luminosity is a critical
luminosity where there exists a balance between gravity and radiative
pressure caused by the young stars formed. When a starburst is tidally
triggered, as is the case in M51, the (starburst) luminosity can grow rapidly.
When this growth happens on a shorter time scale than the dynamical (free
fall) timescale of the starbursting system, in this case giant molecular
clouds, the luminosity grows beyond the critical luminosity and star
formation rate stops. This could explain the narrow burst length observed in
the cluster formation rate. Solomon et al.(1987) give typical values for
molecular clouds in the Galaxy. The dynamical time scale can be estimated
by: $t_{\rm dyn} \simeq R_{\rm eff}/\sigma = R_{\rm eff}^{1/2}$, where
        $\sigma$ is the velocity disperion inside the cloud.  Thus,
        $t_{\rm dyn}$  
        is between 6 - 15 Myr for molecular
clouds in the mass range $10^6 - 10^{7.5} M_{\odot}$ as found in M51
        (Rand \& Kulkarni 1990), which are the most likely progenitors of
        star clusters in M51.  This may explain the short duration of
        the burst in M51.

   \subsection{Spatial Dependence of the Cluster Formation Rate}

\begin{figure}
\begin{center}   \includegraphics[width=9cm]{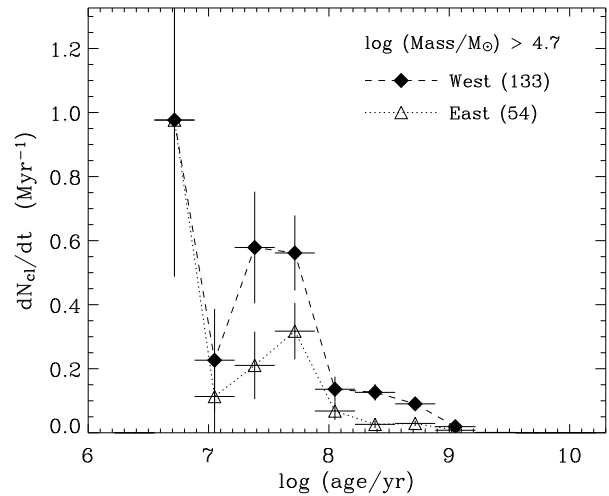}
\end{center}
     \caption{Spatial dependence of the cluster formation rate. We
         have not corrected the rates here for fitting errors.  We note
         that the general trend in the formation rate is similar
         between the two sides of the galaxy.}
         \label{opposite}
   \end{figure}

Figure \ref{opposite} shows that the cluster formation history does not
differ significantly between the two sides of the galaxy, although the
west side of the galaxy seems to have approximately twice as many
clusters as the east side.
This is not an effect of the differing detection limits between the
two pointings, as we have restricted our analysis to magnitudes
brighter than the 90\% completeness limit of the shallowest observation.

The higher cluster formation rate on the west side of the galaxy is
also reflected in the presence of large, isolated and young star
cluster complexes located in the western spiral arms, that appear to be
largely absent in the eastern spiral arms (Bastian \& Gieles 2004).
These complexes are typically less than $\sim 6$ Myr old, contain $3
\times 10^{4}$ to $3 \times 10^{5} M_{\odot}$, and are associated with
large concentrations of CO (Henry et al. 2003). 

Henry et al. (2003) studied the present day star formation rate (SFR)
and its relation to the molecular cloud content in the spiral arms of M51,
based on NICMOS (Pa$\alpha$) and BIMA-SONG (CO 1-0) observations.
They find that both the molecular cloud content and the SFR
are higher in the NW and W  spiral arm (regions 1a and 1b in their
terminology) than in the SE and E arm (regions 2a and 2b).  Thus the
larger amount of clusters that we find in the western arm is
consistent with the measured higher SFR on this side of the galaxy.

   \section{A population of short-lived clusters}
   \label{young-rates}
   As noted in \S~\ref{cfh}, the cluster formation history of M51 shows are
   large increase during that last $\sim 6$ Myr (see
   Fig.~\ref{age1}). 
   This can be interpreted in two ways, either M51 is experiencing a
   burst in cluster formation right now, or that a significant portion
   of the very young clusters ($<$ 10 Myr old) will disrupt within the
   next few Myr.  Three additional observations argue for the latter scenario.
   First, a similar feature has been observed in the NGC 
   4038/39 system (Whitmore 2003).  Second, we know of many
   loose open star clusters within the Galaxy that are expected to
   disperse on timescales shorter than 10 Myr (e.g. Lada \& Lada
   1991).  Lastly, $\sim 10$ Myr is the timescale given for NGC 5253
   for clusters to dissolve and disperse their stars into the field
   (Tremonti et al. 2001).  Adopting 
   this scenario, we can explore what percentage of the young clusters will
   disrupt within the next few Myr.

\begin{figure}
 \begin{center}   \includegraphics[width=8cm]{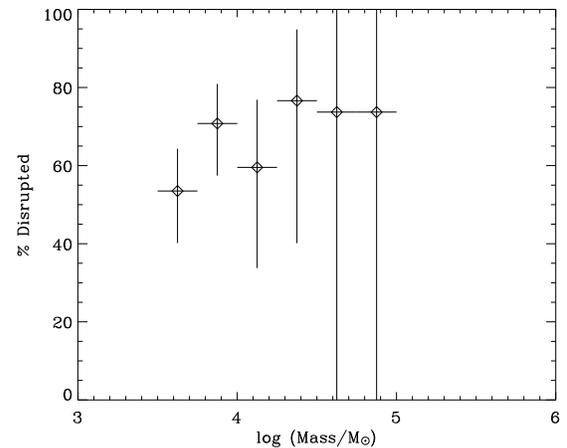}
\end{center}
     \caption{The fraction of the young ($<$ 10 Myr old) clusters per mass
         bin that are expected to disrupt within the next $\sim 10$ Myr.}
         \label{disruption-young}
   \end{figure}

   To do this we compare the observed ``formation'' rates in the first
   two bins of Fig.~\ref{age1}, i.e. the bins of $\log
  (t/{\rm yr})$=6.6 to 6.93, and 6.93 to 7.26.
  In Fig.~\ref{disruption-young}
   we show the percentage of the 
   young clusters that are expected to disrupt before they age to
   older than 10 Myr, as a function of their mass. One striking
   feature of this figure, is that the percentage of the clusters
   disrupted is largely independent of mass within the uncertainties
   of the measurements. 
 
The mean value of the number ratio of clusters in the logarithmic age
bins of (6.93,7.26) compared to (6.6,6.93) for clusters with a mass in
the range of $3.5 < \log (M/M_\odot) < 5.0$ is 68$\pm$ 
15\%. This is comparable to the fraction ($\sim 87$\%) of 
missing clusters less than $10^8$
years (observed relative to predictions based on embedded star
clusters in GMCs) in the solar neighbourhood estimated by Lada \& Lada 
(1991).  This missing fraction is presumably made up of unbound
clusters that have dispersed. 
For clusters to  
disappear from our sample due to expansion, their 
radius should extend over more than $\sim 3$ pixels (with the exact
value depending on the cluster's brightness and the brightness of the
local background), corresponding
to a linear size of 48 pc. For a mean initial velocity dispersion of
the stars in a cluster of about 10 km~s$^{-1}$, this corresponds to a
disruption time of about 5 Myr.

   The largely mass independent destruction rate of the clusters is in stark
   contrast to other observed disruption patterns (Boutloukos \& Lamers
   2003) in which the disruption timescale is heavily dependent on the
   initial mass of the cluster.  Boutloukos \& Lamers (2003) found
   that the disruption time depends on the cluster mass to the power
   $0.6$ for cluster systems in 
   four different galaxies.  Also theoretical disruption studies
   of young cluster populations (e.g. Vesperini 1998, Fall \& Zhang
   2001, Baumgardt \& Makino 2003) have shown a strong dependence of the disruption
   timescale on the mass of the cluster.  This implies that there are two
   distinct stages of disruption. The first, mass independent,
   operates on very short 
   timescales and may be related to the sudden removal of gas from the
   system caused by supernovae and stellar winds (Boily \& Kroupa
   2003).  This effect has been dubbed ``infant mortality'' (e.g. Whitmore
   2004) and
it corresponds to the rapid disruption of unbound clusters.
The second stage of disruption, strongly mass dependent, operates on
   longer timescales and is a combination of different effects, 
   including evaporation, tidal fields, and relaxation.

   Indeed, we expect that the first period of disruption is largely mass
   independent in order that the mass functions (as interpreted
   through the luminosity function) of extremely young ($<$ 10 Myr) cluster
   systems such as in the Antennae (Whitmore et al. 1999) is similar to
   that of intermediate-age populations (300 Myr - 3 Gyr) such as in NGC 7252
   (Miller et al. 1997) and NGC 3610 (Whitmore et al. 2002). 

        \section{Cluster sizes}
        \label{sizes}

	Using the {\tt ishape} algorithm by Larsen (1999), we have attempted
	to derive the effective radius of all clusters on the WF chips in M51
	with an observed F439W magnitude brighter than 22.6 mag.  We have
	restricted ourselves to these brighter clusters, as fainter
	clusters do not have a high enough signal-to-noise ratio for
	unambiguous
	model fits.  Briefly, the algorithm convolves a two dimensional
	analytical model (e.g. a King profile)
	with the PSF and diffusion kernel of the F555W filter for a
	given WFPC2 chip. We have modelled the PSF of the F555W filter
	with {\it Tiny Tim}, at position $x=400, y=400$ for each of
	the three wide field 
	chips.  We 
	have tested whether the fits change significantly if the PSF
	used is from a different portion of the chip, and find that
	the difference is marginal, on the order of 10\% in the
	derived radius. The algorithm then
	compares the model with the observed cluster profile using a
	reduced $\chi^{2}$ test.  For each observed cluster, multiple
	King profiles, all with concentration factor 30, of varying
	sizes were compared and the best fit model was selected.
	We then converted
	the measured FWHM of the profile to effective radius,
	$R_{{\rm eff}}$ (Larsen 1999). 

	Following Larsen (2004), the minimum value accepted for a
	'resolved' cluster was a FWHM of 0.2 pixels.
	At the distance of M51, a FWHM of 0.2 pixels for a King
	profile with concentration factor 30 corresponds to an
	$R_{{\rm eff}}$ of 1.2 pc.  We adopt the more conservative
	lower limit of 2 pc when making the subsample of resolved
	clusters from the full cluster sample. We caution that this
	size criterion biases the sample towards large clusters, but
	due to the resolution of the data, clusters smaller than this
	criterion do not have well determined radii.   This results in 407
	clusters that pass the criteria given in \S~\ref{criteria}, have
	reliable size measurements of $\geq 2$ pc and are located more
	than 3 pixels away from the nearest source.  The final
	criterion was applied in order to avoid contamination by
	neighboring sources in the size determination. The size
	criterion ($R_{\rm eff} \geq 2$ pc) eliminates 552 clusters
	from the sample.  Of the 91 sources
	that were detected in four or more bands, did not pass our
	$\sigma^{2}(BVR)$ criterion, and were brighter than $V =
	22.0$, 84\%
	were unresolved.  This provides further evidence that our
	cluster selection eliminates the majority of contaminating
	individual stars from our sample. 

\begin{figure}
	 \includegraphics[width=8cm]{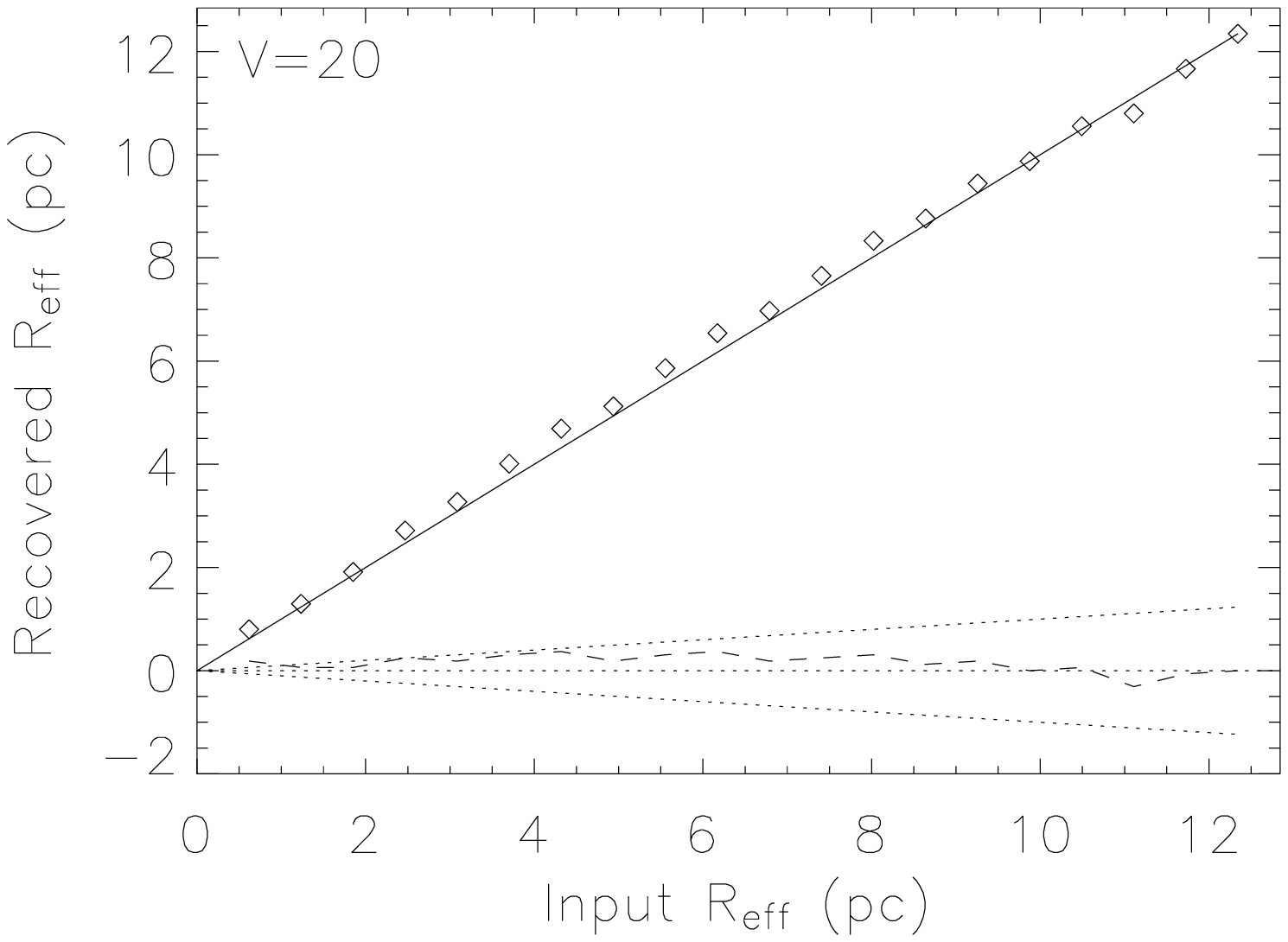}
	 \includegraphics[width=8cm]{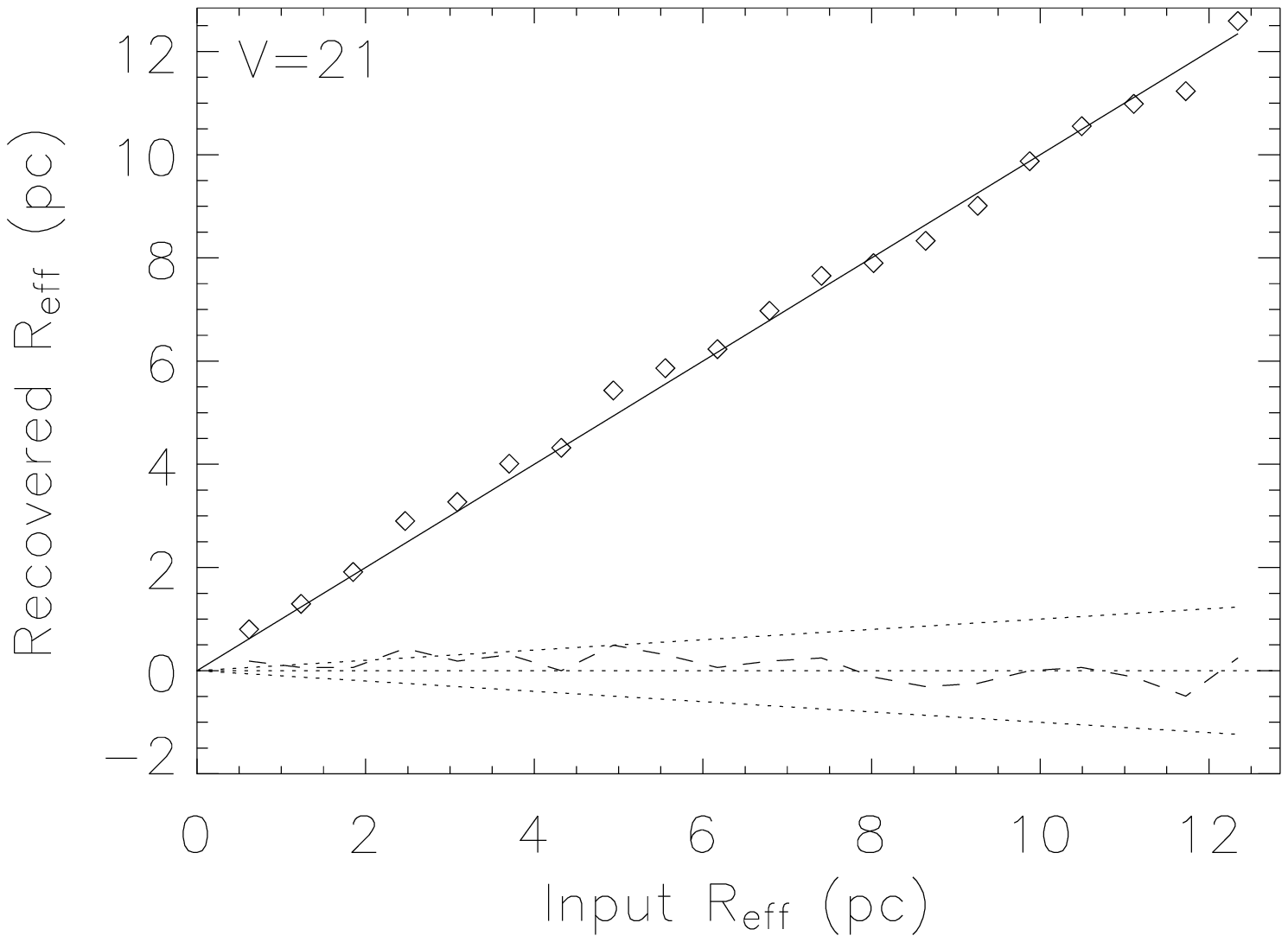}
	 \includegraphics[width=8cm]{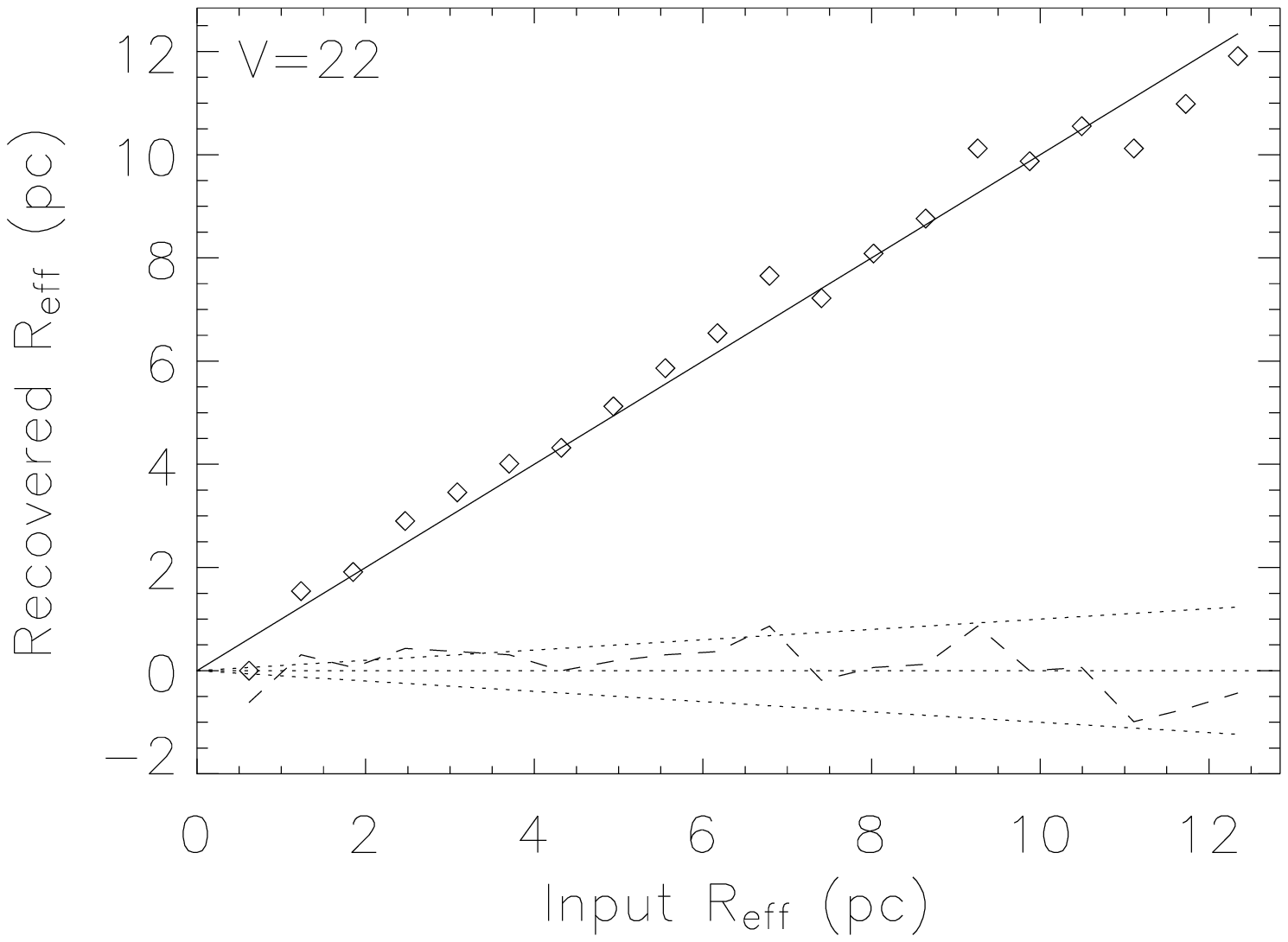}
	 \includegraphics[width=8cm]{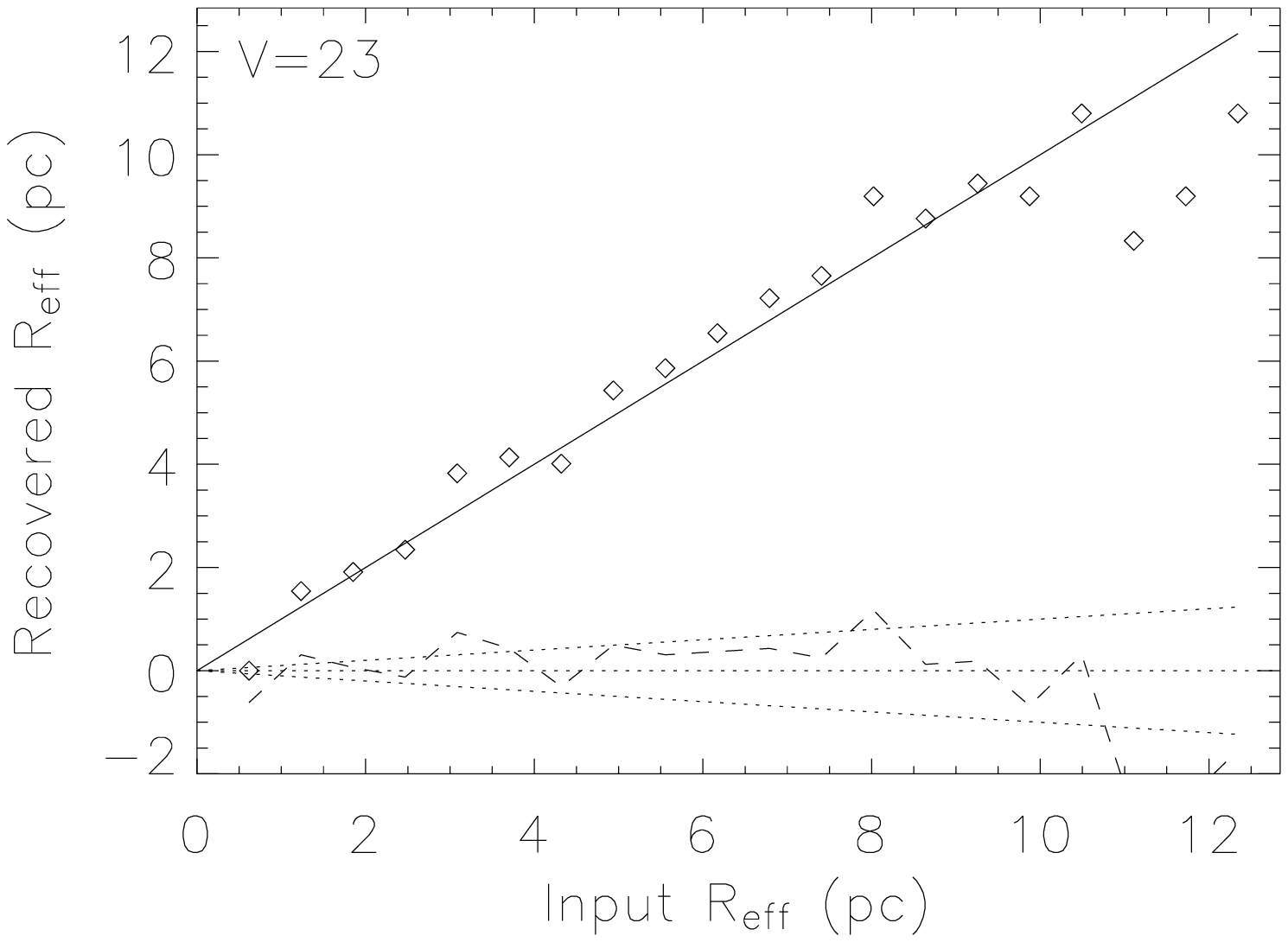}

	      \caption{A test of the accuracy of the measurements of
         cluster radii.  The derived effective radius as a function of
         both the input effective radius and the magnitude.  A
         one-to-one relation is shown by the solid line.  The
         difference between the recovered and the input radii is shown as
         the dashed line, while the dotted lines show the value 10\%
         difference between the input and the recovered radii.}
         \label{size-test}
   \end{figure}

	Additionally, we have tested
	the reliability of the derived fit as a function of the
	magnitude of the cluster.  To do this we generated 20
	artificial clusters with effective radii between 0.65 and 12.4
	pc for magnitudes $V = 20, 21, 22,$ and $23$ mag (i.e., 80 clusters in
	total).  We added these analytic profiles to the F555W
	images and re-fit them using the above method.
	Figure \ref{size-test} shows the results of these tests.  As a
	function of decreasing brightness, the first clusters to
	deviate from the input radius are the large clusters.  This is
	due to the fact that the outer parts of the clusters will be
	lost in the background noise, and 
	so larger clusters will be affected first.

\begin{figure}
   \includegraphics[width=8cm]{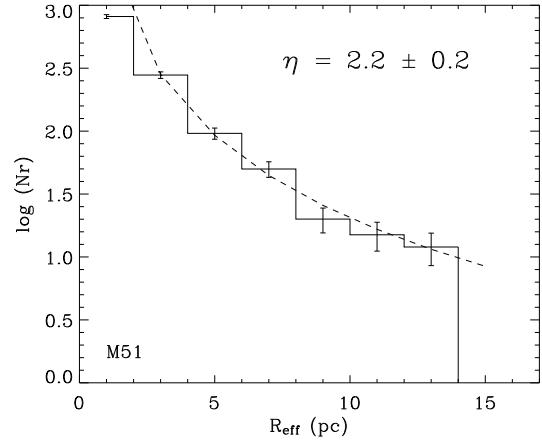}

      \caption{Size distribution of all clusters
   more than 3 pixels away from the nearest source.  The dashed line is
   a power-law  
   fit to the data of the form $N(r){\rm d}r \propto r^{-\eta}{\rm d}r$, with
   $\eta = 2.2 \pm 0.2$.  The fit was carried out for all clusters
   with $ 2.0 < R_{\rm eff} < 15.0$ pc.  The
   power-law fits the data quite well for sources 
   greater than 2 pc, although it over-estimates the number of small
    clusters (unresolved clusters were given a size of 0 pc) with
   respect to the observations.}
         \label{reff}
   \end{figure}

\begin{figure}
   \includegraphics[width=8cm]{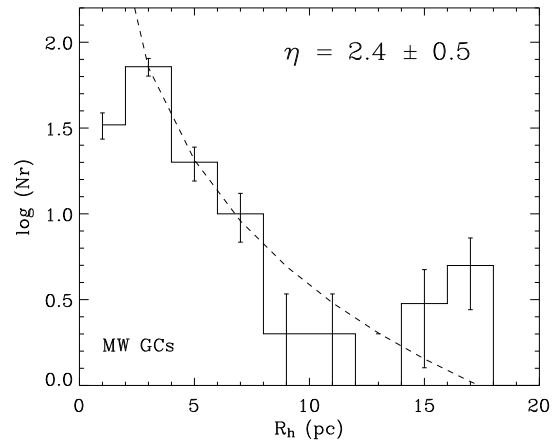}

      \caption{Size distribution (half mass radius) of the Galactic
         globular cluster population. The dashed line is
   a power-law fit to the data of the form $N(r){\rm d}r \propto
         r^{-\eta}{\rm d}r$, with $\eta = 2.4 \pm 0.5$.  The fit was
         carried out for all clusters with $ 2.0 < R_{\rm eff} > 15.0$ pc. }
         \label{gcreff}
   \end{figure}

	 The distribution of sizes for all clusters more than 3 pixels
	 away from the nearest source, is shown in
	 Fig.~\ref{reff}.  The dashed  
	 line in the figure is a power-law fit to the data of the form
	 $N(r){\rm d}r \propto r^{-\eta}{\rm d}r$, with $\eta = 2.2
	 \pm 0.2$, where the fit was carried out only on clusters with
	 $R_{{\rm eff}} > $ 2 pc (i.e. the resolved sample) and
	 $R_{{\rm eff}} < $ 15 pc (most larger sources were determined to
	 be blends of sources upon visual inspection).  We note that
	 this fit overestimates the number of small clusters,
	 a possible indication of a turnover in this distribution.  We compare
	 this to the size distribution of the Galactic globular clusters,
	 by using the data available in the McMaster catalog (Harris
	 1996)\footnote{We used the revised February 2003 verison,
	 http://physun.physics.mcmaster.ca/Globular.html}.
	 Fig.~\ref{gcreff} shows the size distribution 
	 (half-mass radius) of the Galactic globular clusters, along
	 with the best fitting power-law with $\eta = 2.4
	 \pm 0.5$.  We note that the size distribution of the M51
	 clusters is different from that reported for the young
	 cluster system in the merger remnant NGC 3256, which is
	 characterized by $\eta = 3.4$ (Ashman \& Zepf
	 2001).  In that same work, the authors report that the size
	 distribution of the Galactic globular clusters has $\eta =
	 3.4$, significantly steeper than that found here (although
	 using the same catalog).  This may have been caused by the
	 erroneous addition of +1 to $\eta$ in the fitting of the size
	 distribution (see Elmegreen \& Falgarone 1996 for a
	 correction of a similar point of confusion in the data of Galactic
	 GMCs). We therefore report that the size distribution of the
	 old Globular clusters in the Galaxy and that of the young M51
	 clusters (and possibly that of the young NGC 3256 system) has
	 the same form, within the errors.  This distribution is,
	 however, significantly
	 different from the Galactic GMC size distribution, which
	 is characterized by $\eta = 3.3 \pm 0.3$ (Elmegreen \&
	 Falgarone 1996), in contrast with what was reported by Ashman
	 \& Zepf (2001).  

	 The similarity between the size distributions of MW GCs and
	 young star clusters suggests that disruption
	 processes (which have had a much longer time to act on the MW
	 GCs) have not preferentially acted upon the larger clusters.
	 Thus, cluster disruption appears to be largely independent of
	 size.  Additionally, the strong resemblance between size
	 distributions of M51 and the Galactic globular clusters lends
	 support to the idea of a universal formation mechanism for
	 star clusters.

\section{Correlations with the cluster radius}
\label{correlations}

\begin{figure}
   \includegraphics[width=8cm]{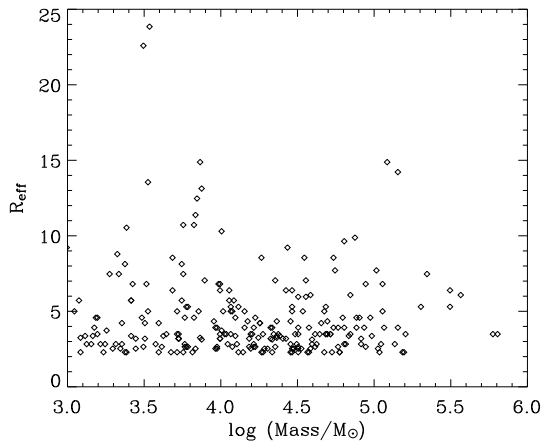}
      \caption{The cluster mass vs. radius for all clusters with
      $R_{\rm eff} > $ 2 pc and located more than 3 pixels
      away from the nearest neighboring source.  }
         \label{mass-vs-reff}
   \end{figure}

With a dataset such as compiled in this work, it is interesting to
search for correlations among the derived parameters.  In particular
we are interested in relations which may reflect the formation
mechanism of star clusters.  Much emphasis has been
placed on the relation between the size of a cluster and the cluster's
mass/luminosity (e.g. Zepf et al. 1999, Ashman \& Zepf 2001).  The
GMCs that are presumed 
to be the birth place of young clusters follow the relation $r_{\rm
cloud} \propto M_{\rm cloud}^{1/2}$ (Larson 1981).  Young clusters,
such as those in merging galaxies (Zepf et al. 1999) and in
spiral galaxies (Larsen 2004) show only a slight correlation (if at
all) between their radii and masses.  The results using our sample are
shown in Fig.~\ref{mass-vs-reff}.  There is no apparent direct
relation between the mass and size of a cluster in M51.

\begin{figure}
   \includegraphics[width=8cm]{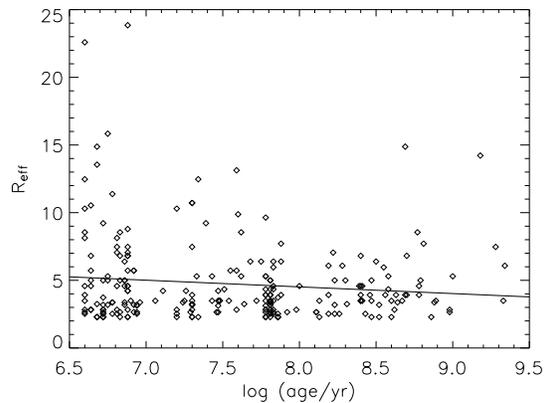}
      \caption{The age of the clusters vs. their effective radii.
   There is a slight tendency for younger clusters to be larger than
   their older counterparts, but the scatter is large.}
         \label{age-vs-reff}
   \end{figure}

 Recently, Larsen (2004) has shown that young clusters in spiral
 galaxies have shallower luminosity profiles than older clusters,
 presumably due to an extended halo surrounding young clusters,
 although no correlation between size and age is found.  In
 \S~\ref{sizes} we presented our method used to derive the cluster sizes,
 which assumed a constant cluster profile, unlike the study of Larsen
 (2004) which also solved for the best fitting cluster
 profile.  Figure ~\ref{age-vs-reff} shows the effective radii vs. ages
 for the resolved clusters in our M51 sample.  No strong correlation is found,
 although there is a slight tendency for younger clusters to be more
 extended than their older counterparts.  We also note that if we
 include the unresolved clusters (assigning an effective radius of 1
 pc) this slight trend is removed.  

 We have also attempted a multi-variable fit of the form:  
\begin{equation}
 R_{\rm eff}=b \times {\rm age}^{x} * {\rm mass}^{y}.
\end{equation}
The derived exponents, $x$ and $y$, are $-0.051 \pm 0.027$ and $0.048 \pm
0.031$, respectively, if we restrict the sample to only the most robust
measurements, $3 < R_{\rm eff} ({\rm pc}) < 10$ and sources more than 5
pixels away from the nearest source.  As before we see a slight trend
of the size 
decreasing with increasing age as well as a slight increase in size
with increasing mass, though both relations are quite shallow and have
rather large uncertainties.

\begin{figure}
   \includegraphics[width=8cm]{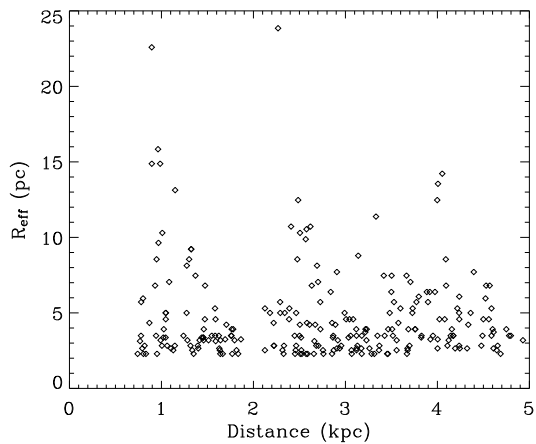}
      \caption{The effective radii of a clusters versus their distance to
   the nucleus of M51.  No trend is apparent in the data.}
       \label{distance-vs-reff}
   \end{figure}

\begin{figure}
   \includegraphics[width=8cm]{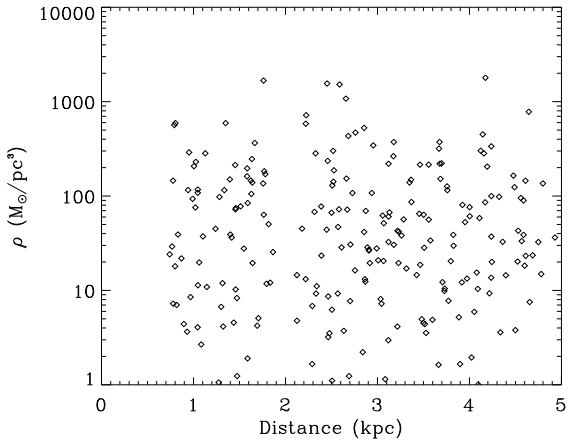}
      \caption{The density of the clusters versus their distance to the nucleus
   of M51.  No trend is apparent in the data.}
         \label{distance-vs-density}
   \end{figure}

   Van den Bergh et al. (1991) report that globular clusters in the
   Galaxy follow the relation $D_{\rm eff}
   \propto R_{\rm gal}^{1/2}$ (with a large scatter) where $D_{\rm
   eff}$ is the diameter 
   within which half the light of the cluster is contained, in projection, and
   $R_{\rm gal}$ is the distance of the cluster to the Galactic
   center.  They interpret this result as suggesting that compact
   clusters form preferentially near the centers of galaxies where the
   density of gas clouds is higher than in the outer regions of the
   galaxy.  We have searched our sample in two ways to see if a similar
   effect is present.  We have looked for a relation between the
   effective radius of a cluster and its galactocentric distance
   (Fig.~\ref{distance-vs-reff}) as well as a trend between the
   stellar density (instead the effective radius) of each cluster and
   its galactocentric distance 
   (Fig.~\ref{distance-vs-density}).  No clear relation can be seen.
   This may mean that the young clusters are forming by a different
   mechanism than globular clusters. This may
   simply be due 
   to our sample coming mainly from the disk of M51 while the globular
   clusters presumably formed outside the Galactic disk.
   Alternatively, the relation found
   for old globular clusters in the Galaxy may be due to disruption
   effects which have had a much longer time to work on globular
   clusters than on the young disk population in the M51 sample.

        \section{Conclusions}
        \label{conclusions}
        We have analysed the star cluster population of the
        interacting galaxy M51.  By comparing the broad-band
        magnitudes with those of SSP models, we have derived the ages,
        extinction values and masses of 1152 clusters with ages
        between 4 Myr and 10 Gyr, and masses between $10^{2.7}
        - 10^{6} M_{\odot}$.  The main conclusions can be summarized
        as follows:
\begin{enumerate}
\item In order to examine the cluster formation history within a galaxy,
        a minimum mass cut-off must be applied, above which the sample
        is complete over the entire age range of interest.  This
        avoids the bias introduced by the effects of the fading of
        clusters due to stellar evolution. 
\item The cluster formation rate increased significantly approximately
        60 Myr ago, which is coeval with the proposed {\it second
        encounter} between M51 and its companion NGC 5195.  We
        interpret this as evidence supporting the multiple encounter
        model of Salo \& Laurikainen (2000).
\item The number of young ($<$ 10 Myr) clusters is much higher than
        would be expected for a constant cluster formation rate over
        the past 20-30 Myr.  By comparing the formation rate during
        the past 10 Myr to that between 10-20 Myr, we find that
        $\sim70$\%  of the young clusters will disrupt before they are
        10 Myr old.  This process appears to be independent of mass,
        which implies that it is a fundamentally different disruption
        mechanism from that which dominates the disruption of older
        clusters.  Thus, there seem to be at least {\it two} distinct
        modes of cluster disruption.  The first mechanism operates
        during the first $\sim$ 10 Myr of the cluster's lifetime while
        the second operates on significantly longer timescales.  We
        suggest that the first may be due to rapid gas removal from the
        emerging young cluster, based on the models of \cite{boily}.
        This is similar to the fraction of unbound clusters
        in the solar neighbourhood.
        The second mechanism has been extensively modelled
        (e.g. Vesperini 1998, Fall \& Zhang 2001, Baumgardt \& Makino 2003) and
        includes two-body 
        encounters within the cluster, disk shocking, and stellar
        evolutionary processes.  This disruption mechanism was studied
        for M51 by Boutloukos and Lamers (2003), based on the cluster
        study in one of the WFPC2 chips presented in Paper I. 
        We will improve this disruption study  in detail in Paper III,
        based on the new and largely extended cluster sample presented
        in this paper.  
\item  The western side of M51 contains a significantly
        higher number of 
        star clusters, although the age and mass distributions of the
        clusters present do not show any large discrepancies.  This
        trend is also reflected in the presence of large, isolated,
        and young star cluster
        complexes located in the west spiral arm, which appear to be
        absent on the east side of the galaxy.  This is also
        consistent with the estimated star formation rates of Henry et
        al. (2003). 
\item  The cluster effective radius distribution can be well
approximated by a power-law of the form $N(r){\rm d}r \propto
r^{-\eta}{\rm d}r$, 
with $\eta = 2.2 \pm 0.2$.  This is remarkably similar to the distribution of
old Milky Way Globular Clusters, which is characterized by  $\eta
\simeq 2.4 \pm 0.5$.  These distributions are significantly different
from the Galactic 
giant molecular cloud (GMC) distribution ($\eta = 3.3$).  The agreement
between the size distributions of the old Galactic globular
clusters and the young clusters in M51 suggests that cluster disruption
is largely independent of cluster size and argues for a universal
cluster formation mechanism. 
\item We have searched our dataset in order to look for correlations
among the derived parameters.  From this study we report that:

\begin{enumerate}
\item  We find a strong correlation between the age of a cluster and
its extinction, with the youngest clusters ($< 10$ Myr) having $<A_V>
\approx 0.55$ mag, while the oldest clusters ($\sim 1$ Gyr) have $<A_V>
\approx 0.30$ mag.  At an age of $\sim 20$ Myr the extinction drops
suddenly, and then remains relatively constant for older ages.  This
is qualitatively expected as clusters emerge 
from the molecular cloud from which they formed.
\item  A shallow relation is found between the effective radius of a
        cluster and its mass.
\item  There is a marginal correlation between the age of a cluster and
its effective radius, with young clusters being slightly larger than
older ones.
\item  Contrary to observations of globular clusters in the Galaxy, no
correlation between the size of a cluster and its galactocentric
radius is found.  Additionally, no correlation between density and
galactocentric radius is apparent in our dataset. 
\end{enumerate}

\end{enumerate}

\begin{acknowledgements}
We thank S\o ren Larsen for useful discussions of the work and
comments on the draft, both before and during the refereeing process.
Additionally, we thank  Michael Fall, Brad Whitmore, and 
Rupali Chandar for useful discussions, and Phillip Massey for help in
estimating the contamination by field stars.  We also thank Cees Bassa
for making Figs. 1 \& 2. 
        
\end{acknowledgements}

\end{document}